\documentclass[a4paper,fleqn,usenatbib]{mnras}

\usepackage{newtxtext,newtxmath}

\usepackage[T1]{fontenc}
\usepackage{ae,aecompl}

\usepackage{graphicx}	% Including figure files
\usepackage{amsmath}	% Advanced maths commands
\title[MCMC Chemical evolution and DustPedia]{A Bayesian chemical evolution model of the DustPedia Galaxy M74}
\author[Calura et al.]
{Francesco Calura$^{1}$, Marco Palla$^{2,1}$, Laura Morselli$^{3}$, Emanuele Spitoni$^{4,5}$, Viviana Casasola$^{6}$,
\newauthor 
Kuldeep Verma$^{7}$, 
Andrea Enia$^{1}$, Massimo Meneghetti$^{1}$, Simone Bianchi$^{8}$, Francesca Pozzi$^{9}$, \newauthor Carlotta Gruppioni$^{1}$
\\ ~ \\
$^{1}$INAF – Osservatorio Astronomico di Bologna, Via Gobetti 93/3, 40129 Bologna, Italy\\
$^{2}$Sterrenkundig Observatorium, Ghent University, Krijgslaan 281 - S9, 9000 Gent, Belgium\\
$^{3}$CINECA National Supercomputing Center, Casalecchio di Reno, Bologna, Italy\\
$^{4}$INAF - Osservatorio Astronomico di Trieste, Via Tiepolo 11, 10134 Trieste, Italy\\
$^{5}$Université Côte d’Azur, Observatoire de la Côte d’Azur, CNRS, Laboratoire Lagrange, France\\
$^{6}$INAF – Istituto di Radioastronomia, Via P. Gobetti 101, 40129, Bologna, Italy\\
$^{7}$Department of Physics, Indian Institute of Technology (BHU), Varanasi-221005, India\\
$^{8}$INAF–Osservatorio Astrofisico di Arcetri, Largo E. Fermi 5, 50125, Firenze, Italy\\
$^{9}$Dipartimento di Fisica e Astronomia, Universitá of Bologna, via Gobetti 93/2, 40129, Bologna, Italy
}
\begin{document}

\label{firstpage}

\pagerange{\pageref{firstpage}--\pageref{lastpage}} \pubyear{2019}

\maketitle

\begin{abstract}
We introduce a new, multi-zone chemical evolution model of
the DustPedia galaxy M74, calibrated by means of MCMC methods.

We take into account the observed stellar and gas density profiles 
and use Bayesian analysis to constrain two fundamental parameters characterising the gas accretion and star formation timescale, i.e. 
the infall timescale $\tau$ and the SF efficiency $\nu$, respectively, as a function of galactocentric radius $R$.
Our analysis supports an infall timescale increasing with $R$ and a star formation efficiency decreasing with $R$, 
thus supporting an 'Inside-Out' formation for M74. 
For both $\tau$ and $\nu$, we find a weaker radial dependence than in the Milky Way. 

We also investigate the dust content of M74, comparing the observed dust density profile with the results of our chemical
evolution models. Various prescriptions have been considered for
two key parameters, i.e. the typical dust accretion timescale $\tau_{0}$ and the
mass of gas cleared out of dust by a supernova remnant, $M_{\rm clear}$, 
regulating the dust growth and destruction rate, respectively.  
Two models with a different current balance between destruction and accretion 
i.e. with an equilibrium and a dominion of accretion over destruction, can equally reproduce the observed dust profile of M74. 
This outlines the degeneracy between these parameters in shaping the 
interstellar dust content in galaxies.
Our methods will be extended to more DustPedia galaxies to shed more light on the relative roles 
of dust production and destruction. 
\end{abstract}

\begin{keywords}
galaxies: individual (NGC 0628)– galaxies: star formation - ISM: dust
\end{keywords}

%%%%%%%%%%%%%%%%%%%%%%%%%%%%%%%%%%%%%%%%%%%%%%%%%% 

%%%%%%%%%%%%%%%%% BODY OF PAPER %%%%%%%%%%%%%%%%%%
\section{Introduction} 
In the last few decades, infrared observations at low and
high redshift opened our view to 
a complete characterisation of the coldest components of galaxies. 
Major results were possible thanks to Spitzer \citep{Werner04} and Herschel \citep{Pilbratt10} space telescopes,
fundamental far-infrared probes that allowed the characterisation and understanding
of the obscured star formation in local and distant galaxies (e. g., \citealt{Rodighiero10,Gruppioni13}). 
Moreover, thanks to the Atacama Large Millimetre/submillimetre Array (ALMA), 
we now know that at high redshift, galaxies had higher molecular gas fractions than local
ones (e.g. \citealt{Tacconi10,Berta13}).  
These results opened a new view to the critical importance of detailed studies of the cold interstellar
medium (ISM), both in the form of neutral and molecular gas. 
In addition to being richer in cold gas and therefore most actively star-forming, we now know also
that galaxies at the peak of star formation were significantly dustier than at the present time.
Infrared (IR) probes enabled the accounting of the majority of star formation in the Universe, emphasising the fundamental 
role played by interstellar dust in galaxy evolution. Beside containing 
fundamental information on the evolutionary path of galaxies, the presence of cosmic dust affects 
a multitude of astrophysical processes, as it is one main contributor 
of the opacity of the gas, crucial for the formation of stars in molecular clouds.
Moreover, dust grains represent efficient catalyst for the formation of molecular hydrogen,
as their surfaces are the ideal environment for such process to occur (\citealt{Gould63,Perets05,Gavilan14}). 
Nonetheless, dust influences the spectral properties of galaxies across an extremely vast range of wavelengths.
The most important effect of dust grains consists in the re-processing of electromagnetic radiation.
This process consists in the absorption of ultraviolet and optical light,
thus making dust the main physical responsible for insterstellar extinction,  
and in its thermal re-emission in the infrared band. 

For a galaxy with average properties like the Milky Way, it has been estimated that up to a half of the radiation emitted by 
its stars is re-processed by interstellar dust \citep{Popescu02,Davies17}. 
Since the short-wavelength light is mostly contributed by young massive stars, the IR emission of dust
is regarded as one major tracer of star formation, in particular in regions where the UV emission is obscured.

Interstellar dust is also an important repository of metals, as it 
can incorporate heavy elements with particular properties into the solid phase, subtracting
them from the interstellar gas phase, in a mechanism called dust depletion.
These elements, called refractory, include abundant chemical species such as Fe, Si, Ca, Mg and C. 
Most of our knowledge of the nature of dust grains comes from Galactic observations and from neaby galaxies,
where accurate measurements of the IR luminosity can be performed and in which the spatial
distribution of dust can be explored in great detail. 
The local Universe offers a great variety of  morphologically
complex, chemically evolved galaxies where key properties such as stellar, dust and gas masses,
chemical abundances, star formation rates can be measured. 

A recent observational program aimed at the comprehensive study of the cold gas and dust
content of local galaxies is represented by the DustPedia\footnote {http://dustpedia.astro.noa.gr/} project (\citealt{Davies17,Clark18,DeVis19,Nersesian19,Casasola20}).  
DustPedia includes all nearby galaxies that have been observed with the Herschel Space Observatory and for which 
multiwavelength observations are available, in a spectral range that extends from the UV to the far-IR. 
The DustPedia sample probes a range of fundamental intrinsic -such as gas, stellar and dust  fractions -
and environmental properties, as it contains galaxies that reside in the field and in clusters \citep{Davies19}. 

One major innovative result 
possible thanks to the richness of this sample concerns the detailed study 
of spatially-resolved scaling relations, such as the galactic Main Sequence
(between SFR and M$_*$, e. g., \citealt{Elbaz07,Noeske07}, \citealt{Popesso19a}, \citealt{Popesso19b}) and the
Schmidt-Kennicutt relation \citep{schmidt59,kennicutt98}, between SFR and molecular gas surface densities, presented in 
\cite{Morselli20}, \cite{Enia20} and \cite{Casasola22}. 

Such broad range of available properties has enabled a series of theoretical studies aimed at using various
approaches to create realistic galaxy evolution models, including numerical simulations.  
Dust has been rarely modelled in simulations, due mostly
to the large number of parameters involved in its production and its uncertainty. 
Despite the significant progress experienced by these models in the last few years 
(\citealt{Mckinnon17,Aoyama18,Gjergo18,Hou19,Granato21}), 
their high computational cost renders the study of the parameter space of dust production and destruction problematic.
The post-process modelling of some properties, like galactic spectra, 
is a valuable alternative, like in the recent comparison of the spectral energy distributions 
and physical properties of simulated and observed galaxies of \cite{Trcka20}. 
One problem with the post-processing approach is that it does not allow a self-consistent modelling of the
main properties of insterstellar dust, such as the evolution of its mass and composition, that by construction cannot be  
 derived directly from the star formation and chemical evolution history of the simulated galaxies. 
Cosmological semi-analytic models for galaxy formation are a viable alternative to hydrodynamical simulations \citep{Baugh06}. 
Some of them include dust production \citep{Valiante11,Popping17,Vijayan19,Triani20,Dayal22} and they allow
a quicker exploration of the paramenter space than hydrodynamic simulations, still within a cosmological context. 
A less computationally expensive approach to theoretical galaxy evolution is represented by chemical evolution models
(\citealt{Tinsley80,Pagel97,Matteucci21}). 
Even if in a non-cosmological context, these tools allow one to trace the evolution of the abundances of a set of chemical elements in a parametric fashion,
in that the star formation and gas accretion history are not modelled directly but parametrised
via a set of semi-analytic recipes. From the computational point of view these models are very light and fast to run, therefore
they allow a rapid and detailed investigation of the parameter space. The results of chemical evolution models have been found to be in good
agreement with the ones of cosmological models, in particular regarding the abundance pattern of stars in the Milky Way \citep{Colavitti08,Calura09b,Spitoni19}.  
Efficient developments related to the use of such models have been possible thanks to their match
with methods for Bayesian analysis. 
Such methods are nowadays 
 used in several areas of astrophysics, including cosmology \citep{Dunkley05}, studies of the propagation of cosmic  rays  \citep{Putze10},
 and of local active  galactic  nuclei \citep{Reynolds12},  Milky  Way  (MW) dwarf  satellites  \citep{Ural15}, 
 semi-analytical models of galaxy formation \citep{Kampakoglou08,Henriques09} 
 and galactic chemical evolution models (e. g., \citealt{Cote16,Spitoni20,DeLooze20}). 
The current applications have been carried on to study mainly the MW, as it represents
the best studied galaxy from the point of view of the properties of the stellar population and chemical abundances
in their main components. 
In this paper, we are interested in the modelling of the spatial distribution of interstellar dust on galactic scales, 
for which an external, nearby galaxy is more suited than the MW. 
The DustPedia galaxy M74 (NGC 0628) is an optimal target for our study, 
since it is observed nearly face-on, hence it permits a very detailed mapping of a few fundamental features, such as the gas,
stellar and dust density profiles \citep{Casasola17}. 
In this paper, we use a Bayesian approach to build a multi-zone chemical evolution 
model tailored to M74. Specifically, our aim is 
to study the spatial distribution and internal variations of the the gas, stellar and dust components of this system. 

This paper is structured as follows. In Sect. 2, we present the main observational properties of the
target galaxy. In Sect 3, we present the basic ingredients of the chemical evolution model used in this study and 
our Bayesian fitting procedure. 
In Section 4, we present our results and a discussion of them,  whereas in Sect. 5, we present our conclusions.

\section{The galaxy NGC 628 (A. K. A. M74)} \label{sec_m74}
The data used in this work have been presented in 
\cite{Morselli20} (M20) and \cite{Enia20} (E20) and are based 
on multiwavelength observations of five nearby face-on spiral galaxies belonging to the DustPedia sample.  
DustPedia is a catalogue  of 875 local systems, built originally with the purpose of studying
the properties of dust in nearby galaxies of various morphological types. 
DustPedia includes every galaxy observed with the 
Herschel instrument within $\sim$30 Mpc distance from the Milky Way and with a diameter $> 1$ arcmin. 
In E20, a coherent sample of galaxies with 
a ‘grand-design spiral’ structure was presented, characterised by Hubble stage index T 
between 2 and 8 (\citealt{deVaucouleurs91,Corwin94}) and a diameter $>$ 6 arcmin. 
NGC 628 (also known as M74) is observed nearly face-on (the inclination is 19.8$^{\circ}$) and at a distance of 10.14 
Mpc (E20). This galaxy has a uniform wavelength coverage, distributed across more than 20 observational bands
and optimal for estimating the physical parameters analysed in this paper, i.e. the stellar, gas and dust surface densities. 

E20 presented a set of multi-wavelength images of M74, ranging from the far-ultraviolet and obtained with the GALaxy Evolution eXplorer
(GALEX; \citealt{Morrissey07}) to the far-infrared (FIR), i.e. at 250 $\mu m$ and obtained
with the Herschel Space Observatory \citep{Pilbratt10}, showing a clear, regular spiral pattern, appreciable across
4 orders of magnitudes in wavelength (see their Fig. 1). 
The physical properties of M74 were determined by 
performing a pixel-by-pixel SED fitting, on 
scales between 0.39 kpc and 1.5 kpc. 
A grid of square cells of 8 arcsec $\times$ 8 arcsec was constructed and, in each cell and at each wavelength, a measure of the flux was derived. A second separation in cells of fixed physical size (1.5 $\times$  1.5 kpc) was also performed for further check of the SED fitting quality and to validate the method (for further details, see E20 and M20). 
The flux in every
available photometric band from UV to far-IR was then obtained,  along with the relative errors and the associated astrometric positions of the cell centres. 
In the SED fitting procedure, only cells with no more than five undetected bands were included. The discarded cells are mostly located in the outermost parts of galaxies. 
SED fitting was performed with the publicly available code MAGPHYS \citep{daCunha08}, optimal for 
panchromatic SED modelling of local galaxies (e.g. \citealt{Viaene14}) and of resolved, sub-galactic regions \citep{Smith18}. 
MAGPHYS allows to simultaneously model the emission observed in the UV-to-FIR regime by assuming that the whole energy output is balanced between the radiation emitted at UV/optical/near-infrared wavelengths and the amount absorbed by dust and then re-emitted in the FIR. 
The physical scales probed in this work (0.39 kpc) are such that the energy-balance criterion holds (E20 and references therein).  
In the energy-balance criterion between stellar and dust emission,
the UV-optical photons emitted by young stars are absorbed (and scattered) by dust grains
in star-forming regions and more diffuse parts of the ISM, and the heated dust then reradiates the absorbed energy in the IR
(e. g., \citealt{Miettinen17,Bianchi18}). 
The energy-balance criterion holds on scales in which the visible stellar component is co-spatial with the obscured,
dust-emitting component of a star-forming region. 
This physical scale, typically of ~200-400 pc (E20; M20) happens to be of the same order
or above the region size where a 'molecular star formation law' can be observationally retrieved, i.e. where 
the scaling between the averaged molecular gas surface density and the star formation rate 
is $\Sigma_{SFR} \propto (\Sigma_{\rm H2})^{\rm N}$, with $N\sim 0.5-1.5$ (e. g., \citealt{Bigiel08,Schruba10,Casasola15}), i.e. where the Schmidt-Kennicutt
law holds \citep{Khoperskov17}. The pixel size of M74 is at the upper edge of this condition,
but still suitable for these purposes. 

By means of a Bayesian approach, the code determines the posterior distribution functions of various physical parameters by fitting the observed photometric data to the modelled emission. The latter is calculated from a set of native libraries composed of 5 $\times 10^4$  stellar population spectra,
with varying star formation histories and derived from the \cite{Bruzual03} spectral library.
Such library allows for a uniform exploration of a few fundamental parameters, such as stellar age, star formation timescale and
metallicity. 
The stellar emission libraries are associated to $5 \times 10^4$ two-component SEDs (\citealt{Charlot00,daCunha08}),
which take into account the emission from stellar birth clouds and the diffuse interstellar medium. 
SED fitting on the integrated  
photometry of the DustPedia data was also performed (E20), as a further validation of the analysis and to derive large-scale properties of the galaxy,
such as total SFR and the gas mass. 
The stellar and dust mass values obtained in each cell were directly taken from the MAGPHYS outputs. 
MAGPHYS models the dust emission as the sum of two components,
the one coming from stellar birth clouds (e.g. PAH, mid-IR continuum and grains in thermal equilibrium) and a colder one from the ambient ISM
\citep{daCunha08}. 

\begin{figure*}
\includegraphics[width=0.41\textwidth]{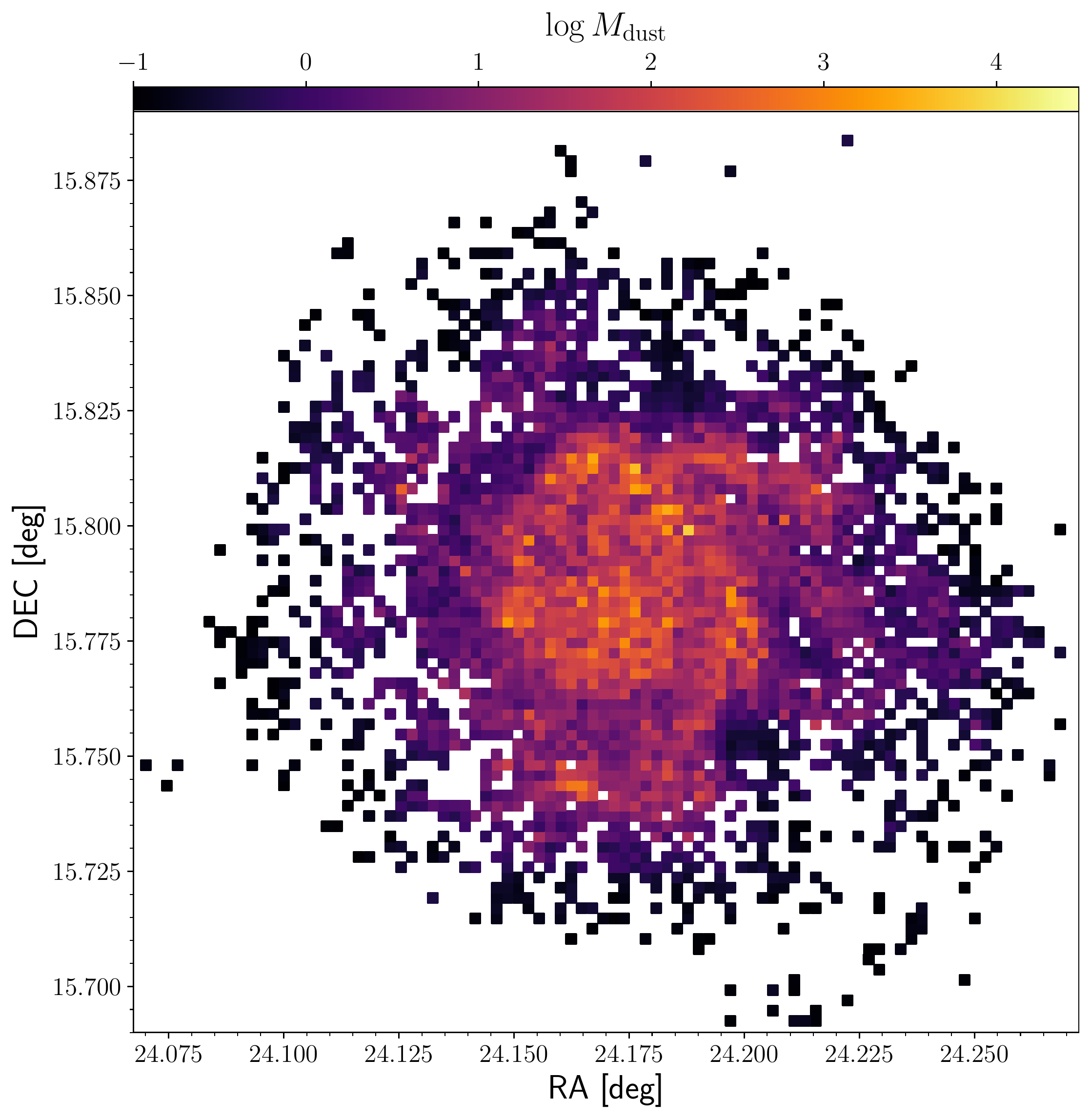}
\includegraphics[width=0.58\textwidth]{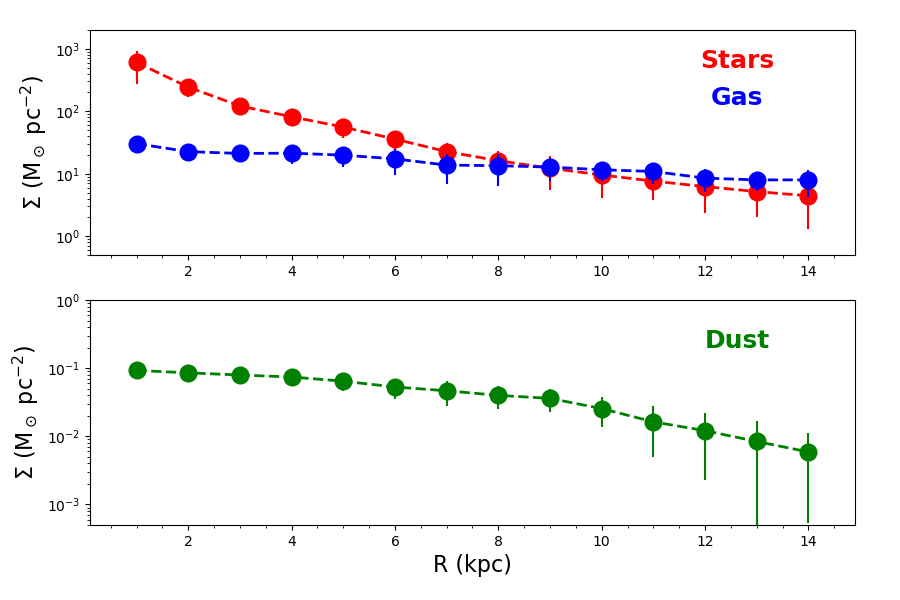}
    \caption{Left: observed dust mass map of M74. In each pixel, the dust mass was derived with the methods described in Sect.~\ref{sec_m74}. 
    Right: observed stellar and gas surface density profiles (upper panel) and dust density profile (bottom panel) of M74. 
    The data used for the calculation of the profiles
    are from \citet{Morselli20} and \citet{Enia20}.}
    \label{fig_tmap}
\end{figure*}

As for the gas mass, in each cell we consider that it is given by the sum of the neutral and molecular gas mass, including also the Helium contribution. 
The neutral hydrogen mass surface densities have been obtained from the 21 cm maps in the framework of the THINGS survey (The HI Nearby Galaxy Survey; \citealt{Walter08}) performed with the Very Large Array (VLA), with a 6 arcsec and 10 arcsec resolution in the robust and natural weighting, respectively. 
The neutral gas surface density has been estimated from equation 5 of \cite{Walter08}.  
The molecular gas surface density has been computed by means of  $^{12}$CO(2-1) intensity maps obtained in the HERACLES survey 
(The HERA CO-Line Extragalactic Survey; \citealt{Leroy09}), 
performed with IRAM 30 with an angular resolution of 11 arcsec. \\
We defer to M20 for further details on the methods used for estimating the neutral and molecular gas mass in M74.   
Fig.~\ref{fig_tmap} shows a dust mass map (left) and estimates of the stellar, gas and dust surface density profiles (right)
of M74.

\section{Model description} \label{sec_model}

The chemical evolution model used in this paper was originally developed to study the abundance pattern 
of the Milky Way 
\citep[e.g.,][]{Matteucci89,Grisoni18,Palla20a}. 
We implicitly assume that the baryonic mass of any spiral galaxy is dominated by a thin disc of stars and gas, in analogy with the Milky Way.
The disc is composed of several independent rings, each 1 kpc-wide, 
without exchange of matter between them.

In chemical evolution models, the time-evolution of the fractional mass of
the element $i$ in the gas within a galaxy and at the galactocentric radius $R$, ${G}_i (R,t)$  is described by the
following equation (here in compact form, for the extended form see, e. g., \citealt{Matteucci12}):
\begin{equation}
\Dot{G}_i (R,t) = -\psi(R,t) X_i(R,t) + \mathcal{R}_i(R,t) + \Dot{G}_{i,\rm inf}(R,t). 
    \label{eq_chem}
\end{equation}
In Eq.~\ref{eq_chem}, $G_i (R,t) = X_i (R,t) G(R,t)$, where $X_i(R,t)$ represents the abundance (in mass) of
the element $i$ (for which, the sum over all elements in the gas is equal to unity) and $G(R,t)$ is the total gas fraction.
The quantity $\psi(R,t)$ is the star formation rate (SFR), namely the 
amount of gas turning into stars per unit time. This quantity is expressed by the 
Schmidt-Kennicutt relation \citep{schmidt59,kennicutt98} law:
\begin{equation}
\psi(R,t) = \nu \Sigma_{\rm gas}(R,t)^K,
    \label{eq_sfr}
\end{equation}
where $K=1.5$, $\Sigma_{\rm gas}$ is the gas surface density and $\nu$ is the star formation efficiency (in Gyr$^{-1}$), 
assumed to be a function of the radius $R$ and parametrised as described later in Sect. \ref{sec:best}. 
The motivation for assuming a radius-dependent SF efficiency is that, as shown in a few previous studies,
it is not possible to reproduce some specific observables of the MW, such as 
the abundance gradients observed in the disc, by assuming a constant SF efficiency (\citealt{Colavitti09,Grisoni18,Palla20a}).

The quantity $\mathcal{R}_i(R,t)$ is the returned fraction of matter in the form of an element $i$ restored by stars into the ISM
through stellar winds, Type Ia and Type II supernova (SN) explosions and computed from the nucleosynthesis prescriptions.  
In this work, we assume for low-intermediate mass stars (LIMS, i.e. stars with initial mass $m < 8$M$_\odot$)
the stellar yields calculated by \citet{vandenHoek97}, whereas for core-collapse and Type Ia supernova (SNe) we adopt the prescriptions of
\citet{Francois04} and \citet{Iwamoto99}, respectively, with a \citet{Scalo86} stellar initial mass function (IMF).
By means of our adopted set of yields it is possible to reproduce the chemical abundance pattern 
of the solar neighbourhood \citep[e.g.,][]{Spitoni19, Vincenzo19,Palla22}.

The third term in Eq. \eqref{eq_chem} is the infall rate, expressed as: 
\begin{equation}
    \Dot{G}_{i,\rm inf}(R,t)=A(R) X_{i,\rm inf}(R)e^{-\frac{t}{\tau(R)}}, 
    \label{eq_inf}    
\end{equation}
where $X_{i,\rm inf}$ is the abundance of the element $i$ in the infalling material, assumed of primordial
composition.
The infall timescale $\tau(R)$ is determined from the available observational constraints of M74 
 by means of the MCMC analysis described in Sect. \ref{sec:best}. 
The quantity $A(R)$ is determined from the observed total surface mass density profile of M74.

In our model, we assume that spiral galaxies do not experience galactic winds. 
In general, galactic winds are known to play an important role in  galaxy evolution and 
the modelling of this process is a very active field (e. g., \citealt{Somerville15,Zhang18}).
A plethora of models have shown that when proto-galaxies are small and are characterised by shallow
potential wells, strong feedback activity resulting from intense star formation can trigger massive outlows,
in which the dense star-forming gas can leave the parent galaxy (\citealt{Naab17} and references therein).
In cosmological simulations, the generation of massive outflows is important 
to avoid the overproduction of stars, to remove low angular momentum-material and thus for producing realistic discs 
(\citealt{Brook11,Marinacci14,Murante15,Vogelsberger20} and references therein). 

Also observations of local discs show that they still have an important role, and that
they can be described by biconical flows of multiphase gas and dust perpendicular to the galaxy \citep{Rupke18}. 

Our assumption of neglecting the effects of winds in the long-term evolution of a massive galactic disc
is supported by other models that include the evolution of the infall and outflow rates of MW-like disc galaxies,
showing that the latter are generally sub-dominant with respect to the former \citep{Toyouchi15}.
Furthermore, our assumption is strengthened by the fact that a large fraction of the material ejected through outflows
is later reaccreted. 
This is supported by cosmological simulations of galactic discs of different mass, 
which confirm that the majority of the mass ejected by the disc in galaxies with virial mass $>10^{11}~M_{\odot}$ (encompassing the mass of M74)
is reaccreted (\citealt{Christensen18}, see also \citealt{Oppenheimer08,Dave11,Hopkins23}). 
Note that our non-cosmological models for galactic discs do not exclude that winds may have an important role in their low-mass progenitors at early times,
as shown by the history of the Milky Way and of its building blocks \citep{Vincenzo19}.

\subsection{The best model of M74 through Bayesian analysis}
\label{sec:best}
The best-fit chemical evolution model of M74 has been obtained by means of a Bayesian analysis based on Markov Chain Monte Carlo
(MCMC) methods. 
We briefly recall our main assumptions and refer the
reader to \citet{Spitoni20} for a more detailed description of
the fitting method.

In the context of parameter estimation, Bayes' theorem enables the calculation 
of the posterior probability distribution (PPD) of the parameters, as expressed by 
\begin{equation}
 P({\bf \Theta}|{\bf x})=\frac{ P({\bf \Theta})}{ P({\bf x})}P({\bf x}|{\bf \Theta}),
 \label{eq:bayes}
\end{equation}
where ${\bf x}$ represents the set of observables, ${\bf \Theta}$ the set of model parameters,
$P({\bf x}|{\bf \Theta}) \equiv L$ the likelihood, defined
as the probability of observing the data given the model parameters.  
In this work, the set of observables is represented by 
the stellar and gas surface density profiles at the present time, respectively $\Sigma_{\rm \star}$  and $\Sigma_{\rm gas}$, 
computed over $N=14$ radial bins (see the right panel of Fig. \ref{fig_tmap}), i.e. 
${\bf x} =  \{ \Sigma_{\rm \star}, \ \Sigma_{\rm gas} \}$.

In  this study, we statistically constrain the behaviour of the gas accretion timescale and the the star formation efficiency with galactocentric
radius. 
The assumptions of radial dependent infall timescale and SF efficiency have been widely adopted in previous chemical evolution models.  
In the framework of the inside-out model for the formation of the Milky Way, 
\citet{Matteucci89} introduced a radius-dependent time-scale for the exponential infall rate of the Galactic disc, 
increasing linearly as a function of the galactocentric radius  (see also \citealt{Romano00}).
On the other hand, a variable SF efficiency as a function of galactocentric radius was introduced by \citet{Colavitti09} as an additional 
mechanism to explain the abundance gradients in the MW disc (see also \citealt{Spitoni15,Palla20a}).

In our approach, we consider a set of 4 free parameters ${\bf \Theta} = \{A, \ B, \ C, \ D\}$,
in the following two equations:  
\begin{equation}
\tau {\rm   [Gyr]}=A \cdot R +B   \; 
\label{eq_tau} 
\end{equation}
and 
\begin{equation}
\nu {\rm[Gyr}^{-1}{\rm]} =\frac{C}{R} +D \; .
\label{eq_nu} 
\end{equation}

To define the likelihood in eq.~\ref{eq:bayes}, we assume that the uncertainties on the 
observables are normally distributed. In this case, the logarithm of the likelihood can be
written as: 
\begin{equation}
\ln  L  =-\sum_{n=1}^N\ln\left(\left(2 \pi \right)^{d/2} \prod_{j=1}^d \sigma_{n,j}\right)
-\frac{1}{2}\sum_{n=1}^N\sum_{j=1}^d\left(\frac{x_{n,j}-\Sigma_{n,j}}{\sigma_{n,j}} \right)^2,
\label{eq:likelihood}
\end{equation}
where $N=14$ is the number of radial bins, 
$x_{n,j}$ and $\sigma_{n,j}$ are the measured value and associated uncertainty, respectively, of the $j-$th observable 
and in the $n-$th radial bin, whereas 
$\Sigma_{n,j}$ is the theoretical $j-$th mass surface density (i.e. for the gas or stars) at the present day in the $n-$th radial bin, with $d=2$.  

The calculation of the PPD defined in eq.~\ref{eq:bayes} requires the 
specification of the priors on the 4 model parameters.
In our sampling of the PPD, for all the parameters we consider a flat prior in the range (-50,+50); 
only combinations that lead to positive $\tau$ and $\nu$ are admitted. 
The affine invariant MCMC ensemble sampler, "\textit{emcee}: the mcmc hammer" code\footnote{\href{https://emcee.readthedocs.io/en/stable/}{https://emcee.readthedocs.io/en/stable/};
\href{https://github.com/dfm/emcee}{https://github.com/dfm/emcee}}  (\citealt{goodman2010,foreman}) has been used to sample the PPD.

\begin{figure*}
    \centering
    \includegraphics[width=8.2cm]{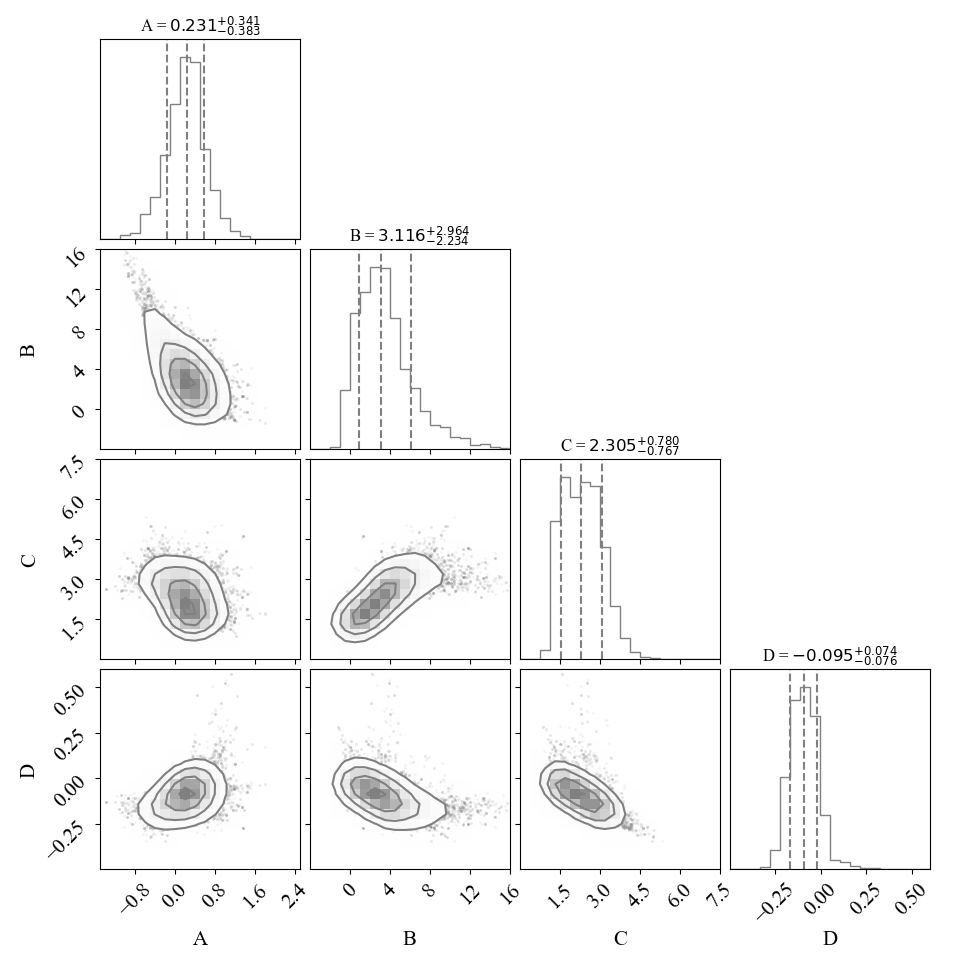} 
    \includegraphics[width=9.2cm]{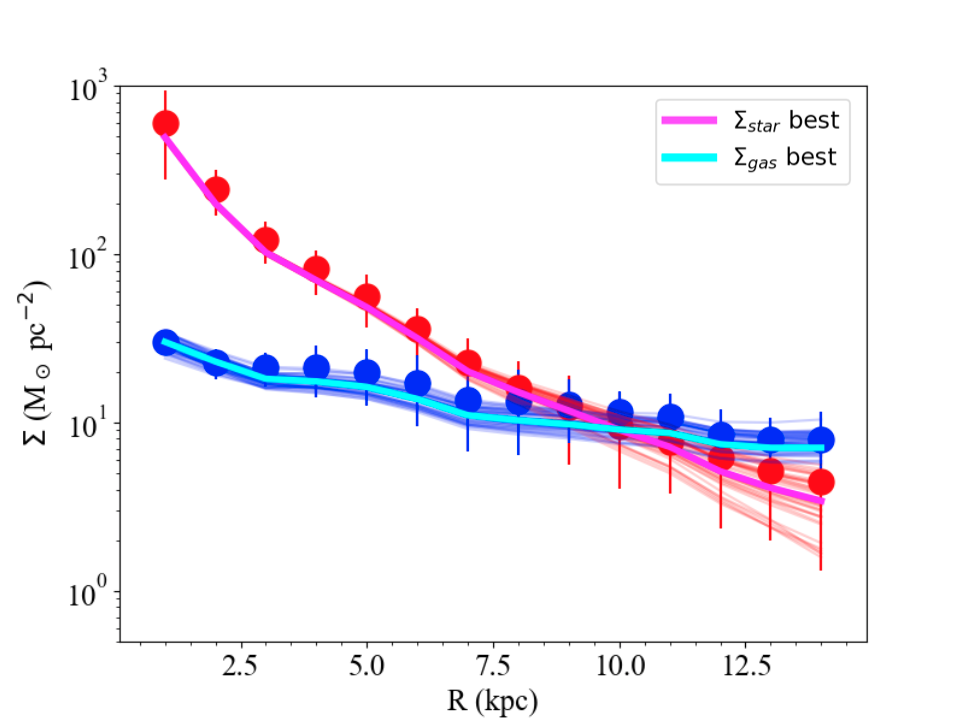}
\caption{Left panel: posterior probability density function of the parameters used to model the accretion and SF history of M74. 
The best-fit value obtained for the parameter (defined in Sect.~\ref{sec3}) 
is reported on the top of each column, along with the associated uncertainties. 
The results of the best-fit model are shown in the right panel, where the thick magenta and cyan lines represent the
best-fit stellar and gas surface density as a function of radius, respectively. 
The blue and red thin lines represent the range of models
obtained from the Monte Carlo walkers at the last step that are 
consistent with the observed profiles, 
i.e. characterised by $\Sigma_{\rm gas}$ and  $\Sigma_{\rm *}$ values within the observed error bars.}
    \label{fig_best} 
\end{figure*}

\subsection{Dust evolution prescriptions}
\label{s:dust_model}

In this work, we adopt the same formalism used in previous works which included the chemical evolution of dust in galaxies of various morphological
types \citep{Dwek98,Calura08,Gioannini17b,Palla20c}.  
The equation used to determine the amount of the chemical element $i$ contained into dust grains is: 
\begin{equation}
\dot{G}_{i,d}(R,t)= - \psi(t) X_{i,d}(R,t) + R_{i,d}(R,t) + \frac{G_{i,d}}{\tau_{i,accr}}(R,t) -\frac{G_{i,d}}{\tau_{i,destr}}(R,t) 
\label{e:dust_evo}
\end{equation}
Similarly to Eq. \eqref{eq_chem},  $G_{i,\rm d} (R,t) = X_{i, \rm d} (R,t) G(R,t)$, where $X_{i, \rm d} (R,t)$
represents the abundance (in mass) of the element $i$ present in dust grains.
The first term on the right-hand side of Eq. \eqref{e:dust_evo} accounts for the amount of dust destroyed by astration, 
i.e. incorporated into new stars. 
$R_{i, \rm d}(R,t)$ accounts for stellar dust production, which in our model is from asymptotic giant branch (AGB) stars and core-collapse SNe. 
In low and intermediate mass stars (LIMS, i.e. stars with mass $0.8 \le M_{\rm \odot}\le 8 $), dust is
produced during the AGB phase. 
For AGB stars, in this work we use the metallicity-dependent dust yields of \citet{DellAgli17}.
Stars more massive than $8 M_{\rm \odot}$ end their life as core-collapse (CC) SNe and are the main producers of dust.
For these sources, we assume the dust yields of \citet{Bianchi07} that, as shown in \cite{Gioannini17}, provide 
dust mass values intermediate between other prescriptions from the literature. 

The third term on the right side of Eq. \eqref{e:dust_evo} accounts for dust accretion (or growth) in molecular clouds. 
In this process, occurring in the coldest and densest regions of the ISM, pre-existing seeds of dust can grow in size as due to accretion
of further refractory elements on their surface (e. g., \citealt{Savage96,Hirashita11,Inoue11,Asano13,Martinez21,Konstantopoulou22}),
enabling a significant increase of the global dust mass 
(e.g. \citealt{Dwek98}; \citealt{Asano13}; \citealt{Mancini15}; \citealt{DeVis17b}). 
The dust accretion rate is regulated by the accretion timescale $\tau_{\rm accr}$.  
For a given element $i$, this quantity can be expressed as:
\begin{equation}
\tau_{i,\rm accr}=\frac{\tau_0}{(1 - f_i)},
\label{eq:tau}
\end{equation}
where $f_i=G_{i,\rm d}(t)/G_i(t)$ is the dust-to-gas ratio for the element $i$ at the time $t$ and $\tau_0$ is the characteristic dust growth timescale.
This formalism was first introduced by \cite{Dwek98}, in which 
the key quantity $\tau_0$ was assumed constant and generally of the same order of magnitude of the typical molecular cloud lifetime
(see also \citealt{Calura08}). 
Later, more elaborate assumptions have been proposed by other authors, with a dependency on fundamental galactic parameters such as metallicity 
and cold gas fraction (e. g., \citealt{Asano13}).  

The fourth term of Eq. \eqref{e:dust_evo} accounts for dust destruction, occurring in the warm and hot phases of the ISM
mostly as due to exploding SNe. 
In this process, the refractory material incorporated in dust grains and accelerated by SN shocks 
is cycled back into the gas phase via various mechanisms
(evaporation in grain-grain collisions, sputtering, thermal sublimation, desorption, see \citealt{McKee89}, \citealt{Jones96})

Similarly to dust accretion, dust destruction is expressed in terms of a grain destruction timescale $\tau_{\rm destr}$, which can be written as:
\begin{equation}
\tau_{i,\rm destr}=\frac{G_i(t)}{(\epsilon \, M_{\rm swept}) SN_{\rm rate}},
\label{eq_taudestr}
\end{equation}
where $SN_{\rm rate}$ is the sum of the CC and Type Ia SN rates, $M_{\rm swept}$ is the ISM mass swept by a SN shock and $\epsilon$ is
the efficiency of grain destruction in the ISM.\\
The product of the two latter terms represents the ISM mass completely cleared out of dust by a SN: 
\begin{equation}
M_{\rm clear}=\epsilon\, M_{\rm swept}. 
\end{equation} 
In this paper, we will study in detail 
the roles of the main quantities regulating dust growth and destruction, i.e. $\tau_0$ and $M_{\rm clear}$ and see how they determine 
the dust content of M74, as traced by the observed dust mass profile as a function of galactic radius. 

The Bayesian approach requires a specific parameterization of a given process 
as a function of fundamental quantities that regulate them, such as the infall timescale and the SF efficiency as a function of the galactocentric
radius (see Equations ~\ref{eq_tau} and ~\ref{eq_nu}).

In this work, we want to test various prescriptions for quantities regulating both destruction and accretion,
in which the dependence on some fundamental physical variables is diversified.  
Testing every single prescription by means of a MCMC would be very challenging  
from a computational point of view; therefore, in this work we choose to model the evolution of dust without a Bayesian approach. 
In the forthcoming future, the next step will be to adopt a Bayesian method to treat also the evolution of dust.

%%%%%%%%%%%%%%%%%%%%%%%%%%%%%% RESULTS %%%%%%%%%%%%%%%%%%%%%
\section{Results} \label{sec3}
In this section we present the results of the best-fit
chemical evolution model of M74, obtained from the observational
constraints, i.e. the observed gas and stellar
density profiles of M74, and by means of the Bayesian method described in Sect.~\ref{sec:best}. 
The best model is then used to study the star formation history and the
properties of the dust in our target galaxy, 
in order to derive constraints on a few fundamental prescriptions adopted in our 
model.

We have used 100 walkers to sample the parameter space. The chains have been run for 500 steps.
The MCMC reaches convergence in less than 125 steps, used as a warm up phase
(burn-in). Therefore, the PPD has been built using the final 375 steps.
As an additional check, we also verified that the acceptance fraction of walkers is about 0.4, well within the range of values recommended (between 0.2 and 0.5).

The posterior probability density function of the parameters is shown in the left panel of Fig. \ref{fig_best}, where 
the best-fit values obtained for our set of parameters are also reported with their uncertainties. 

For each parameter, we compute the best-fit value, the lower and upper uncertainty from 
the median, the 16-th and the 84-th percentile of the distribution (reported on the top of each column of Fig. \ref{fig_best}), respectively. 

We show the results of the best-fit model in the right panel of Fig. \ref{fig_best} (thick magenta and cyan lines), compared to the observed profiles.
The thin lines in Fig. \ref{fig_best} represent the range of models
within the 16th and 84th percentiles, hence corresponding approximately to $1-\sigma$ error bars of the best-fit models. 

\subsection{Star formation and infall timescale as a function of radius}
\label{sec_best_res}

\begin{figure} 
    \vspace{0.25cm}
    \includegraphics[width=1.025\columnwidth]{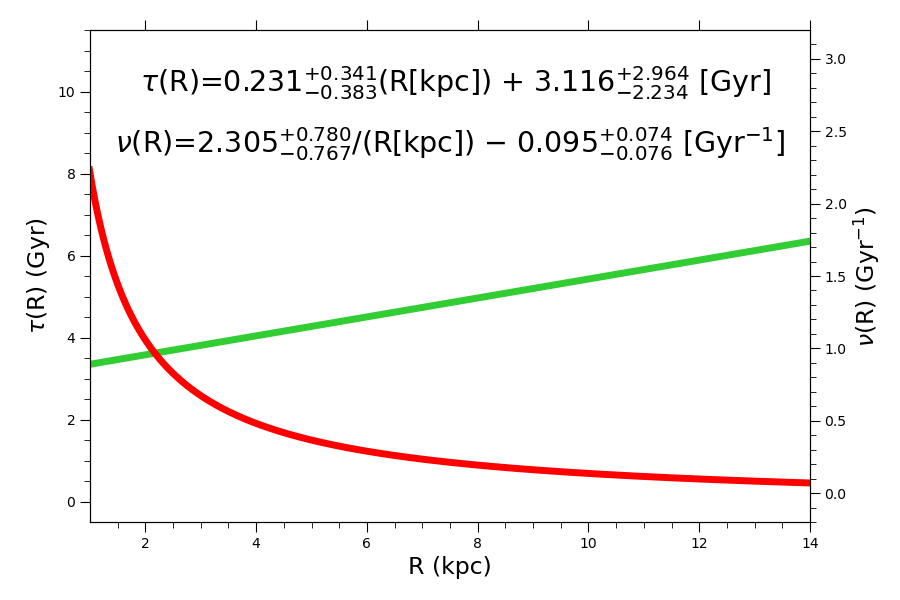}
    \caption{Accretion timescale $\tau$ (green solid line and left-side y-axis) and SF efficiency $\nu$ (red solid line and y-axis on the right)
as a function of galactocentric radius $R$ of the best model obtained for M74 through our Bayesian analysis.}
    \label{fig_laws}
\end{figure}

Fig. ~\ref{fig_laws} shows the infall timescale $\tau$ (green solid line) and the SF efficiency $\nu$ (red solid line)
as a function of galactocentric radius $R$ of the best model obtained for M74 through our Bayesian analysis. 
As for the best-fit $\tau-R$ function (green line), an analogue relation was found 
for our Galaxy by \cite{Romano00}, expressed by
\begin{equation}
\tau_{\rm D, MW} = 1.03~$\rm R$ - 1.267 ~\rm Gyr
\label{eq_tauD_MW}
\end{equation}
and describing the 'Inside-Out' formation model of the disc component of the MW.

The relation in Eq.~\ref{eq_tauD_MW} is able to account for various constraints, i.e. 
the 'G-dwarf' metallicity distribution observed in the solar neighbourhood and in other regions of the Galactic disc,
the gas density profile and the Galactic abundance gradients (see also \citealt{Chiappini01}).

Also for M74, our analysis supports 
an inside-out formation scenario, but the flatter shape of the $\tau-R$ function obtained here suggests 
more similar gas accretion rates at different radii with respect to the MW. 
For comparison, the accretion timescales of M74 at
2 kpc and 14 kpc are $\sim 3.8$ Gyr and $\sim 6$ Gyr, respectively, whereas in the Milky Way, at the same radii 
\cite{Romano00} derived accretion timescales of 1 Gyr and 13 Gyr, respectively. 

In principle, beside a smaller average stellar age, 
a flatter accretion timescale with radius is expected to produce a lower dispersion of
stellar age than in the Milky Way. 
It is however worth stressing that the single-infall model used here differs by construction from
the one used for our Galaxy, which allows for a better characterisation of the various sub-components, 
for which much more observational constraints are available. 
In particular, valuable constraints in this context are represented by the stellar chemical
abundance pattern, at the moment available for our Galaxy and for a few of its satellites only, 
which provide an insightful picture of the past evolution of such systems and therefore
allows one to have tighter constraints for their accretion history.

The red line in Fig. \ref{fig_laws} shows the SF efficiency $\nu(R)$ of M74 derived in this work.
Also in this case, it is interesting to compare this quantity to the analog relation found in the MW. 
\cite{Palla20a} used the two-infall model to constrain some fundamental properties
of the thin and thick Galactic discs, including the SFR profile and the abundance gradients,
considering different chemical elements (including Mg, Fe, N, O). 
Again, their model is more complex than ours since, besides including two
separate infall events for different sub-components, they also take into account radial gas flows.
A variable efficiency of star formation is considered for the thin disc and the effects of other parameters are investigated,
including the delay between the formation of the two discs. 
They find that the assumption of a variable SF efficiency along the disc plays a fundamental role
in reproducing the observed abundance gradients (see also \citealt{Grisoni18}).
Although the trend is similar, i.e. characterised by decreasing $\nu$ at increasing radii, 
the $\nu-R$ relation derived here, which spans a range between $\sim~2$ Gyr$^{-1}$ in the innermost regions
and $\sim 0.1$ Gyr$^{-1}$ in the outskirts, 
is less extreme than the one found in the MW, which ranges between $\sim$5 Gyr$^{-1}$ and $\sim 0.1$ Gyr$^{-1}$.

As additional check, we have compared the metallicity gradient obtained with the best model of 
M74 to a set of available observations. 

\begin{figure}
    \includegraphics[width=0.975\columnwidth]{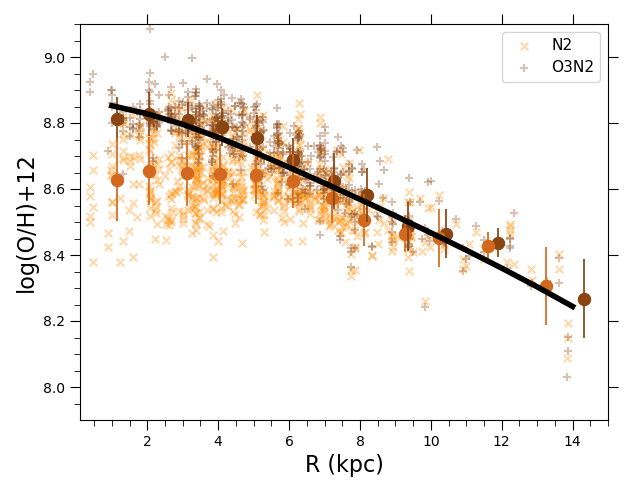}
    \caption{Metallicity gradient of the best model of M74 (black solid line) compared to a set of available observations.
    The metallicity tracer is oxygen, for which two strong line calibrations (the N2 [orange crosses] and the O3N2 [brown crosses])
    have been used to derive the abundances as a function of radius (see \citealt{DeVis19}, \citealt{Bianchi22}).
    The brown and orange circles represent the average abundances computed  
    at different radial bins and considering both sets of abundances.
    In each radial bin, the brown vertical bar associated to the computed
    average abundance represents the $1-\sigma$ uncertainty.}  
    \label{fig:gradient}
\end{figure}

The used metallicity tracer is oxygen. We show in Fig. \ref{fig:gradient}
the data from the DustPedia compilation of \cite{DeVis19}, obtained 
using two different metallicity calibrators: the simplest N2 calibrator, which is based on the ratio between  [NII]$\lambda$6584 and the adiacent
H$\alpha$ line (thus depending little on dust extinction), and the O3N2 method, that uses [OIII]$\lambda$5007, [NII]$\lambda$6584,
H$\beta$ and H$\alpha$ lines \citep{Pettini04}. 
It is well known that, when using different diagnostics, the metallicity measures are significantly sensible
to the adopted calibration \citep{Kewley08,Calura09,Maiolino19}. 
The difference between the two profiles is useful to envisage the uncertainties in the calibration schemes. 
As shown in Fig. \ref{fig:gradient}, our model can satisfactorily account for the observed metallicity gradient, in particular 
the one measured from the O3N2 line ratio, across the entire radial range of the observations (from R=1 kpc to R$\sim14$ kpc). 
As for the gradient measured from N2, in the innermost regions ($R<5$ kpc) 
the data indicate a flatter gradient than what found in our best model. 
Investigating the reason of the 
discrepancy between the N2 measures and the model results is beyond the scope of the present work. 

In Fig. \ref{fig_evol} we show the evolution of the gas mass, stellar mass and star formation history of the best model, obtained as
described in this section. We also show the evolution of the dust mass of the two best models described in Sect. \ref{sec_asa} and
\ref{sec_mat}.

In Table \ref{table_m74}, we present a few observational features of M74 and compare them to the results of our best models. 
The observed and theoretical gas, stellar and dust masses have been computed integrating the respective profiles.
M74 has an observed dust mass of 2$\times~10^7$ M$_{\rm \odot}$, similar to the of other spiral galaxies of similar stellar mass \citep{Pastrav20}.
The results of the comparison between the measured dust mass and the models will be discussed in the Sect. ~\ref{s:dust}.

\begin{table*}
  \begin{center}
\small
  \caption{Summary of the main observational properties of M74 and as computed by means of our best models.} 
  \begin{tabular}{l|ccccccc}
  \hline
                 &   D         & D$_{25}$     & Hubble type     &   M$_{\rm gas}$   &   M$_{\rm *}$                   &  SFR                    &   M$_{\rm dust}$    \\
                 &   (Mpc)     & (')            &               &  (M$_{\rm\odot}$)   &  (M$_{\rm\odot}$)                 &  (M$_{\rm\odot}$ yr$^{-1}$)  &   (M$_{\rm\odot}$)      \\     
\hline
 M74 (observed)  &    9.0$^1$      &  10.0$^2$   &    Sc$^3$       &   8.5$\times~10^9$   &     2.0$\times~10^{10}$     &    0.7$^4$     &    2.0$\times~10^{7}$    \\
 Model (5')      &    -        &   -            &                 &   7.0 $\times~10^9$  &   1.68$\times~10^{10}$       &   0.9    &    1.96$\times~10^{7}$                            \\
 Model (9')      &    -        &   -            &                 &   7.0 $\times~10^9$  &  1.68$\times~10^{10}$        &   0.9    &    2.26$\times~10^{7}$           \\
\hline	      
\end{tabular}

\begin{tabular}{l}
1: \cite{Casasola17}\\
2: D$_{25}$ is defined as the length corresponding to the projected major axis of a galaxy at the isophotal level of 25 mag arcsec$^{-2}$ in the B-band \citep{Casasola17}. \\
At the distance of M74, this corresponds to a diameter of 26.18 kpc. \\
3: From the NED database.\\
4: \citet{Gusev14}
\label{table_m74}
\end{tabular}

\end{center}
\end{table*}

\subsection{Effects of radial flows}
\label{sec_radf}
Radial gas flows may take place as a dynamical consequence of infall \citep{Spitoni11,Bilitewski12}.  
The cosmological accretion of gas is expected to occur mostly in the outskirts of galactic discs \citep{Ho19}. 
The infalling gas has a lower angular momentum than the gas orbiting within the disc, 
and the mixing between the two is expected to induce a net, inward-directed radial flow \citep{Mayor81},
with a velocity in the range of a few km s$^{-1}$ \citep{Lacey85}. 
Radial flows may also originate as a consequence of the viscosity of the gas \citep{Thon98} 
and of spiral density waves that may lead to large-scale shocks and energy dissipation 
(\citealt{Spitoni11} and references therein). 

In our reference multi-zone model we have neglected radial flows,
i.e. we have assumed that no exchange of matter can occur between two adjacent rings.
This assumption relies upon the results of various studies, who have shown that, under reasonable conditions, radial flows
alone have a marginal role in reproducing the abundance gradients measured in the MW disc
(\citealt{Portinari00,Spitoni11,Palla20a}), and that the latter are  
the result of the interplay of various processes, including a star formation threshold,
an efficiency of star formation and infall law varying with radius \citep{Colavitti09,Palla20a}.

The study of radial gas flows in local spiral galaxies is a recent, growing field \citep{DiTeodoro21} 
but their effects on the properties of other galaxies have rarely been addressed.  
In this Section, we include radial gas flows in the model of M74 and show their effects on the observables considered
in this work, i.e. the stellar and gas density profiles and the metallicity gradient.

The model used here is described in \cite{Spitoni11}, \cite{Mott13}, \cite{Palla20a} and is based on the formalism of \cite{Portinari00}. 
It is convenient to define the $k-$th shell of the galactic disc 
in terms of the galactocentric radius $R_{k}$, whose inner and outer edges are
$R_{k-1/2}$ and $R_{k+1/2}$, respectively. 
Through these edges, gas inflow can occur with velocity $v_{k-\frac{1}{2}}$ and $v_{k+\frac{1}{2}}$ , respectively,
with velocities assumed to be positive outward and negative inward. 

The net radial flow rate at galactocentric radius $R_{k}$ and time $t$ 
is included as a further additive term in Eq.~\ref{eq_chem} and is expressed as: 
\begin{equation} 
 \Dot{G}_{i, \rm rf}(R_{k},t)= -\, \beta_k \,
G_i(R_k,t) + \gamma_k \, G_i(R_{k+1},t), 
\end{equation}

where $\beta_k$ and $\gamma_k  $ are, respectively:

\begin{equation}
\beta_k =  - \, \frac{2}{R_k + \frac {R_{k-1} + R_{k+1}}{2}}  
	\times \left[ v_{k-\frac{1}{2}}
	    \frac{R_{k-1}+R_k}{R_{k+1}-R_{k-1}}  \right]  
\end{equation}

\begin{equation}
\gamma_k =  - \frac{2}{R_k + \frac {R_{k-1} + R_{k+1}}{2}} 
	     \left[ v_{k+\frac{1}{2}}
	     \frac{R_k+R_{k+1}}{R_{k+1}-R_{k-1}} \right] 
	     \frac{\Sigma_{(k+1), \rm T}}{\Sigma_{k, \rm T}}.
\end{equation}

$\Sigma_{(k+1),\rm T}$ and $\Sigma_{k, \rm T}$ are the total present-day density profiles 
at the radius $R_{k+1}$ and $R_{k}$, respectively. 

In this parametrization the most critical quantity is the
radial velocity pattern, expressed by $v_{k}$.
We have studied the effects of an inward-directed radial gas flow, 
testing various values for $v_{k}$ and various assumptions for the radial dependence of the infall and SF law.

As for the velocity pattern, we have considered both constant and radial-dependent values,
previosly considered to model radial flows in the Milky Way (\citealt{Grisoni18,Palla20a}). 
As for the infall timescale and star formation efficiency radial dependence, beside the ones of
best-fit model described in Sect.~\ref{sec_best_res}, we have tested other assumptions, including constant $\tau$ and $\nu$  (i.e. no
radial dependence) and other, intermediate choices between constant and the ones of the best-fit model.

As for the best-fit model obtained through Bayesian analysis with added radial flows, 
none of the tested assumptions for the velocity pattern 
could reproduce the considered constraints simultaneously. 
The results of one of these cases is shown 
in the left panels of Fig.  \ref{fig_radialf}, where constant radial velocity $v_{k}=-0.5$ km s$^{-1}$ was assumed. 
This is a low value in the framework of models generally suited to account for the velocity
pattern of the MW disc (e. g., \citealt{Palla20a}).
Although it does not have any major effect on the present-day gas and stellar mass profiles but a slight steepening, 
such a low value is enough to steepen significantly the metallicty gradient, as shown in the bottom-left panel of ~\ref{fig_radialf}.
We have verified that other velocity patterns assumed for the MW disc, including a higher (in modulus) constant velocity $v_{k}=-1$ km s$^{-1}$ and a variable
velocity $v_{k}= -(\frac{R_k}{4} - 1)$ km s$^{-1}$ \citep{Grisoni18} cause a steepening also of the density profiles and do not improve the
fit to the observed metallicity gradient. 

In the right panels of  ~\ref{fig_radialf}, we show the results of a couple of models with different assumptions regarding 
the radial dependence of  $\tau$ and $\nu$. 
One model ('Constant $\tau(R)$ and $\nu(R)$', solid lines), is with constant $\tau=4$ Gyr and $\nu$=0.45 Gyr$^{-1}$  at all radii. 
These values for $\tau$ and $\nu$ are similar to the ones of the best model obtained with the MCMC at 4 kpc. 

Another model ('Intermediate  $\tau(R)$ and $\nu(R)$', dashed lines in the right panels of Fig.~\ref{fig_radialf}) is characterised by values for the quantities $A$ and $C$ defined in Sect.~\ref{sec:best} that are approximately half of the best-fit model,
but values for $\tau(R)$ and $\nu(R)$ at $R=4$ kpc that are comparable to the ones of this model.
These choices are certainly arbitrary; however,  more tests have been performed by anchoring the relations to the $\tau$ and $\nu$ of
the best-fit model at other values of $R$, without any improvement in the results. 

\begin{figure*} 
    \vspace{0.25cm}
\includegraphics[width=13.2cm,height=10.2cm]{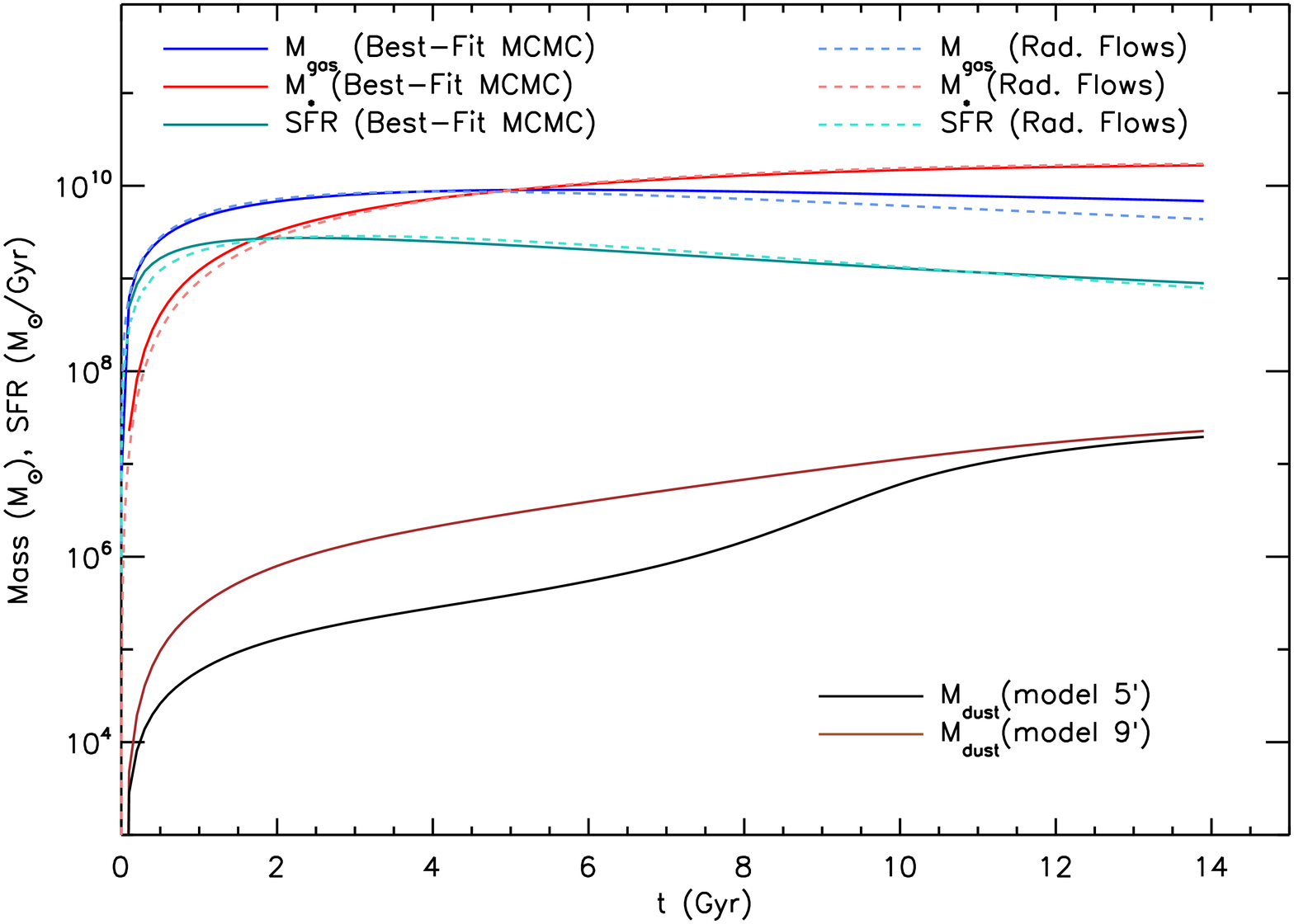} 
    \caption{Evolution of the gas mass (blue solid line), stellar mass (red solid line) and star formation history (dark-cyan solid line)
    of the best-fit model obtained with our bayesian analysis 
described in Sect.~\ref{sec:best}. 
 The light-blue, coral and turquoise dashed lines are the evolution of the gas mass, stellar mass and star formation history, respectively,
 of the 'Intermediate  $\tau(R)$ and $\nu(R)$' model, which includes radial flows and is described in Sect.~\ref{sec_radf}.
The black and brown solid lines are the dust mass of the models 5' and 9' described in Sect. \ref{sec_asa} and
\ref{sec_mat}, respectively.}
    \label{fig_evol}
\end{figure*}

\begin{figure*} 
    \vspace{0.25cm}
    \includegraphics[width=11.2cm,height=8.2cm]{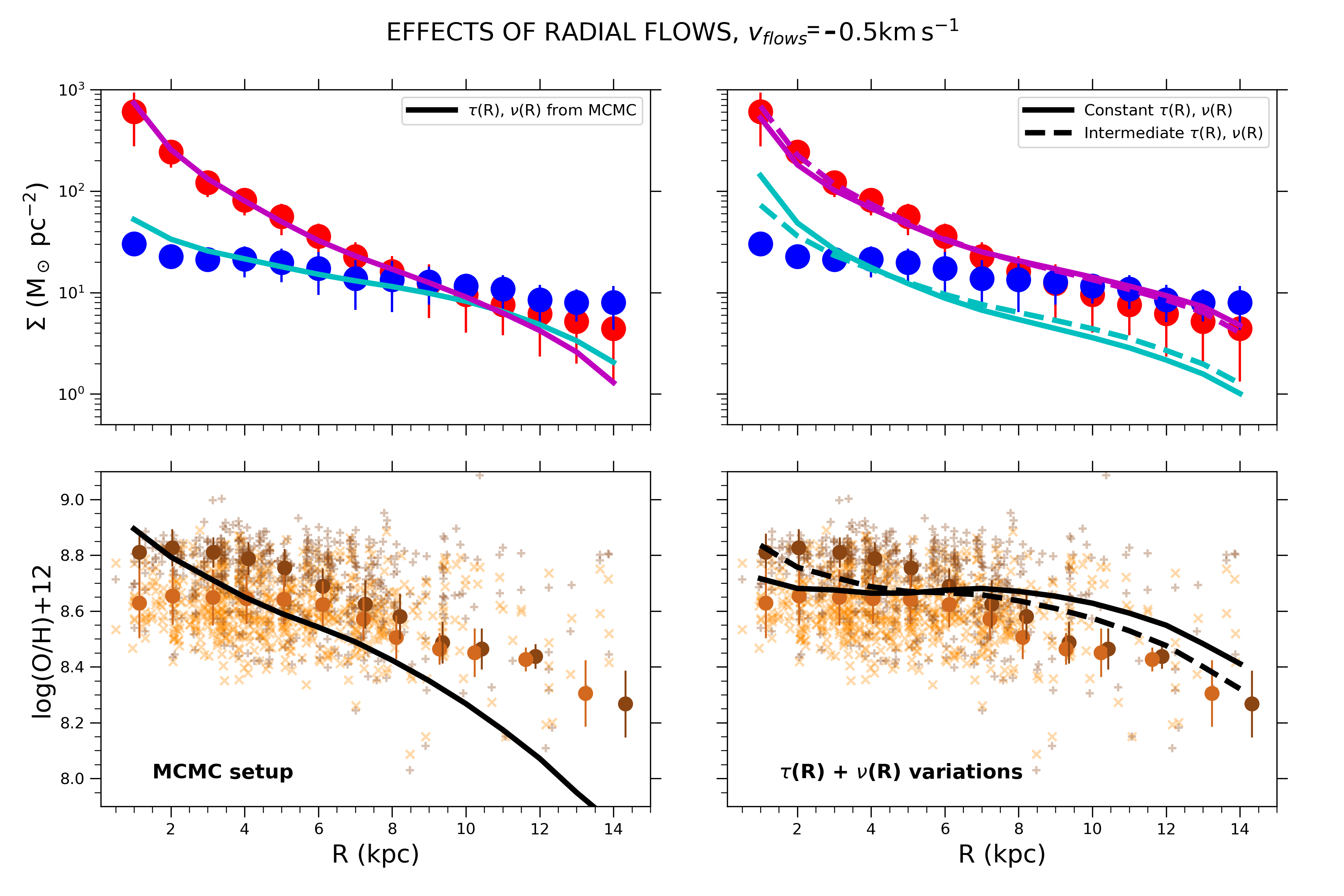}
    \caption{Top-left panel: stellar (solid magenta line) and gas (solid dark cyan line) surface density as a function of radius of the
best-fit model for M74 obtained with the bayesian analysis plus radial flows, described in Sect. \ref{sec_radf}. The blue and red solid circles
are as in Fig. \ref{fig_tmap}. 
Bottom-left panel: metallicity gradient of the best-fit bayesian model of M74 plus radial flows (black solid line) compared to a set of observations 
(other symbols as in Fig. \ref{fig:gradient}). 
Top-right panel: stellar (magenta lines) and gas (dark cyan lines) surface density profiles of the model with constant $\tau(R)$ and $\nu(R)$ 
(solid lines) and intermediate  $\tau(R)$ and $\nu(R)$ (dashed lines), both including radial flows and described in Sect. \ref{sec_radf}. 
The blue and red solid circles are as in Fig. \ref{fig_tmap}. 
Bottom-right panel: metallicity gradient of the model with constant $\tau(R)$ and $\nu(R)$ (solid line)
and intermediate  $\tau(R)$ and $\nu(R)$ (dashed line). 
Other symbols are as in Fig. \ref{fig:gradient}.}
    \label{fig_radialf}
\end{figure*}

Our results are useful to illustrate the overall effects of an inward-directed velocity pattern. 
These models provide a stellar profile in agreement with data and within the error bars at almost all radii,
an acceptable fit to metallicity gradient but a gas density profile steeper than the available observations.
As shown in Fig.~\ref{fig_evol}, the overall effects on the star formation history are modest.

Although our results can certainly be improved with a more thorough 
investigation of the parameter space, the analysis carried out in this 
Section highlights that, by assuming significant net radial inflows in 
the galaxy, we are not able to reproduce the different observational 
constraints of M74, i.e. the stellar density, gas density and 
metallicity gradient simultaneously. The same conclusion also holds by
varying the prescriptions for the radial dependence of gas accretion and SF law. 

At present and based on extant data, it is not possible to firmly assess the radial velocity pattern of the gas in local galaxies.   
A systematic study of gas flows in local spirals based on atomic hydrogen emission line high-sensitivity data 
indicates that the magnitude of radial inflow motion seems small, of the order of
a few 0.1 M$_{\odot}$/yr and a factor 5-10 smaller than the typical SFR values \citep{DiTeodoro21}.
This suggests that radial inflows alone are not strong enough to sustain star formation in local spiral galaxies. 

\subsection{Dust density profile}
\label{s:dust} 

Understanding the roles of destruction and accretion in regulating the dust mass budget in galaxies represents one major
challenge in galaxy evolution studies. 
How these mechanisms depend on key galactic properties, such as radius and SFR, is not known, and various authors
have proposed different pescriptions to include their effects in galactic evolution models. 

In the following, we exploit the best model obtained for M74 to 
test various prescriptions for these two processes and compare our results with one key observational constraint 
available for this galaxy, i.e. the  dust surface density profile. 

In particular, we test different formalisms for the 
parameter regulating the accretion timescale $\tau_0$ of Eq. \ref{eq:tau}  and 
for the destruction timescale, expressed by the 'cleared mass' parameter $M_{\rm clear}$ of Eq.~\ref{eq_taudestr}. 
A summary of the models considered in this work and adopting 
different prescriptions for accretion and destruction is presented
in Table  \ref{tab_models}.

%%%%%%%%%%%%%%%%%%%%%%%%%%%%%%%%%%%%%%%%%%%%%%%%%%%%%%%%%%%%%%%%%
\begin{table*}
  \begin{center}
  \caption{Models investigated in this work. 
First column: model number; second column: adopted stardust prescriptions; third column: dust destruction model;
fourth column: adopted dust accretion prescriptions.}
  \label{tab_models}
  \begin{tabular}{cccc}
\hline
MODEL    & Stardust prescriptions         & Destruction          &  Accretion      \\
\hline
1        &   D17 (AGB), B07 (SNe)         & \cite{Asano13} ($M_{\rm clear}$: Eq. \ref{eq_mclear_asa13})       &   \cite{Asano13} ($X_{\rm cold}=0.05$)  \\

2        &   D17 (AGB), B07 (SNe)         & \cite{Asano13} ($M_{\rm clear}$: Eq. \ref{eq_mclear_asa13})       &   \cite{Asano13} ($X_{\rm cold}=0.25$)\\

3        &   D17 (AGB), B07 (SNe)         & \cite{Asano13} ($M_{\rm clear}$: Eq. \ref{eq_mclear_asa13})       &   \cite{Asano13} (variable $X_{\rm cold}$)  \\

4        &   D17 (AGB), B07 (SNe)         & \cite{Priestley22} ($M_{\rm clear}$: Eq.  \ref{eq:Priestley})   &   \cite{Asano13} ($X_{\rm cold}=0.25$)  \\

5        &   D17 (AGB), B07 (SNe)         & \cite{Temim15} ($M_{\rm clear}$: 4000 $M_{\rm \odot}$) &   \cite{Asano13} ($X_{\rm cold}=0.25$)  \\
\hspace{.05cm}   5'       &   D17 (AGB), B07 (SNe)         & \cite{Temim15}              &   \cite{Asano13} ($X_{\rm cold}=0.25$)  \\
                          &                                & ($M_{\rm clear}$: 2667 $M_{\rm \odot}$ at R= 1 kpc; 4000 $M_{\rm \odot}$  at R$>1$ kpc)            & \\                             
\hline
6        &   D17 (AGB), B07 (SNe)         & \cite{Asano13} ($M_{\rm clear}$: Eq. \ref{eq_mclear_asa13})       &   \cite{Mattsson12} ($\xi=500$)  \\
7        &   D17 (AGB), B07 (SNe)         & \cite{Priestley22} ($M_{\rm clear}$: Eq.  \ref{eq:Priestley})     &   \cite{Mattsson12} ($\xi=500$)  \\
8        &   D17 (AGB), B07 (SNe)         & \cite{Asano13} ($M_{\rm clear}$: Eq. \ref{eq_mclear_asa13})       &   \cite{Mattsson12} ($\xi=225 \nu$)   \\
9        &   D17 (AGB), B07 (SNe)         & \cite{Asano13} ($M_{\rm clear}$: Eq. \ref{eq_mclear_asa13})       &   \cite{Mattsson12} ($\xi=150/\nu$)  \\
\hspace{.05cm}   9'       &   D17 (AGB), B07 (SNe)         & \cite{Asano13}              &   \cite{Mattsson12} ($\xi=150/\nu$)  \\
                          &                                & ($M_{\rm clear}$: 2/3 of Eq. \ref{eq_mclear_asa13} at R=1 kpc; Eq. \ref{eq_mclear_asa13} at R$>1$ kpc)            & \\                             
\hline	      
\end{tabular} 
\end{center}
\end{table*}
\label{tab_models}
%%%%%%%%%%%%%%%%%%%%%%%%%%%%%%%%%%%%%%%%%%%%%%%%%%%%%%%%%%%%%%%%%%%%%%%%%%%

\begin{figure*}
    \includegraphics[width=13.2cm,height=6.cm]{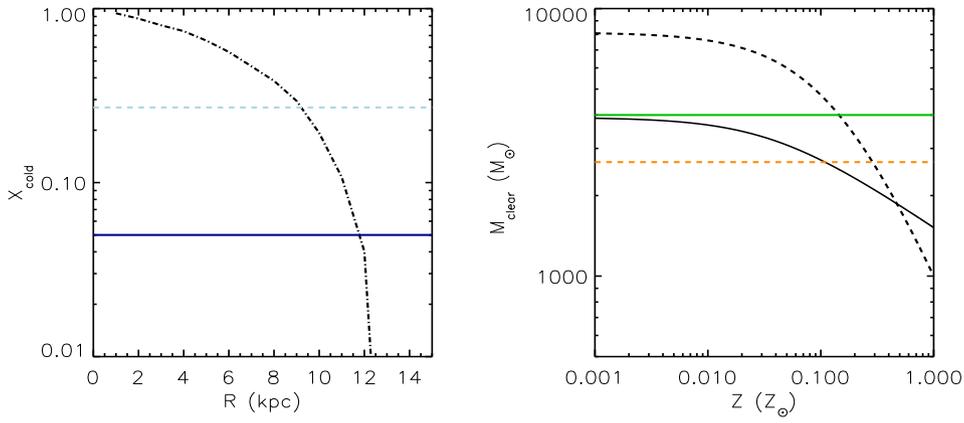}
    \caption{Summary of the adopted prescriptions for dust accretion and destruction.
    Left panel: cold gas fraction $X_{\rm cold}$ as a function of galactocentric radius as in Eq. \ref{eq_xcold}
    (black dash-dotted line)
    and constant, with a value of 0.05 (blue solid line) and 0.25 (light-blue dashed line).
    Right panel: interstellar mass cleared out of dust $M_{\rm clear}$ from \citet{Asano13} (black solid line), from
    \citet{Priestley22} (black dashed line) 
    and constant with a value  4000 M$_{\rm \odot}$ (green solid line) and of 2667 M$_{\rm \odot}$ (corresponding to 2/3 or the previous value, 
    orange dashed line) as adopted 
    in the innermost regions (see Sect.~\ref{sec_mat}).}  
    \label{fig:dustpresc}
\end{figure*}

\subsubsection{Asano et al. (2013) prescriptions for dust accretion}
\label{sec_asa}
\citet{Asano13} investigated  the role of
dust mass growth in the ISM by means of a closed-box chemical evolution model which included
stardust production, destruction by SN shocks and accretion. 
They emphasised the role of the interstellar metallicity in determining the dust accretion timescale,
identifying a critical metallicity value, decreasing with decreasing SF efficiency,  
above which accretion becomes the primary process 
in regulating the galactic dust budget, also causing the strong depletion of heavy elements from the gas phase.

In the model of \citet{Asano13}, the dust accretion timescale $\tau_0$ of Eq. \ref{eq:tau} is expressed as: 
\begin{equation}
    \tau_0= \frac{\tau_{\rm g}}{X_{\rm cold}} 
\end{equation}
where $\tau_g$ is the characteristic dust growth timescale within a molecular cloud and $X_{\rm cold}$ 
is the mass fraction of dense cold clouds in a galaxy. 
In the complete equation of \citet{Asano13} for the characteristic timescale $\tau_g$, this quantity can be expressed
as a function of grain size $a$, sticking coefficient of refractory elements $\alpha$ and mean velocity of metals in the gas  $\langle v \rangle$  
as 
\begin{equation}
\tau_g = \frac{4 \langle a\rangle^3 \rho_{\rm d}}{3 \langle a\rangle^2 \alpha \rho_{\rm ISM} Z \langle v\rangle},
\label{eq_tg}
\end{equation}
where $\langle a \rangle ^2$ and $\langle a \rangle^3$ are the second and third moment of the grain size distribution, respectively, 
while $\rho_{\rm d}$ and  $\rho_{\rm ISM}$ are the average dust and ISM mass density. % $\rho=$
\cite{Asano13} showed that assuming for the typical molecular cloud temperature a value of 50 K,
100 cm$^{-3}$ for the cloud ambient density and an average value of 0.1 $\mu m$ for the grain
size, it is possible to express $\tau_g$ as:
\begin{equation}
\tau_g =  2.0\cdot 10^7 \times \bigg( \frac{Z}{0.02} \bigg)^{-1} \, [{\rm yr}],
\label{eq_asa_tg} 
\end{equation} 
The cold mass fraction $X_{\rm cold}$ is highly uncertain and plays a crucial role in shaping the dust density profiles; therefore,
we will test various assumptions for its value. For this quantity  we test two constant values, embracing a reasonable range supported by observations and
arguments related to the local cold gas budget. The adoption a the low value for $X_{\rm cold}$ 
is supported by estimates of the dense gas mass fraction in molecular clouds in the Milky Way,
performed through observations of the 
1.1 mm dust continuum emission (\citealt{Battisti14}). In this work, observations of 
high column density subregions 
indicate mass fractions as low as 0.07, to be regarded as upper limits to the true mass fractions. 
Motivated by these results, we assume for our model a value of $X_{\rm cold}=0.05$ (Model 1 of Tab.~\ref{tab_models}). 

A larger value for $X_{\rm cold}$ can be derived instead from estimates of
the local budget of neutral and molecular gas.
For the local molecular and neutral comoving gas density, 
\cite{Fukugita04} 
report estimates $\Omega_{\rm H2}=1.6 \times 10^{-4}$ and $\Omega_{\rm HI}= 4.2 \times 10^{-4}$ (expressed as density parameters),
respectively.
From the ratio between the two, we obtain $X_{\rm cold}\sim 0.25$ (Model 2). 

As third option, for $X_{\rm cold}$ we assume the profile derived for M74 as a function
of the galactocentric radius (Model 3), expressed as
\begin{equation}
X_{\rm cold} (R) = \frac{M_{\rm H2} (R)}{M_{\rm H2}(R) + M_{\rm HI}(R)},
\label{eq_xcold}
\end{equation}
 where $M_{\rm HI} (R)$ and $M_{\rm H2}(R)$ are the mass of HI and H2 observed in M74, respectively,
at the radius R.
The adopted prescriptions for dust accretion are shown in the left panel of Fig. \ref{fig:dustpresc},
whereas in the right panel we show the prescriptions for dust destruction. 

As for the dust destruction, the most crucial parameter is the ISM mass cleared out of dust that, in the formulation of
\cite{Asano13}, can be expressed as: 
\begin{equation}
   M_{\rm clear} = 1535 n_{\rm ISM}^{-0.2} \cdot (Z/Z_\odot + 0.039) ^{-0.289}\,[{\rm M}_\odot],
\label{eq_mclear_asa13}   
\end{equation} 
and depends on the ISM particle density $n_{\rm ISM}$ and on the metallicity of the gas $Z$.
The idea behind a metal-dependent cleared mass is that gas cooling is a function of metallicity,
in that  metal-poor gas cools less efficiently, thus remaining
at high temperatures for longer, increasing the amount 
of dust destroyed by thermal sputtering in shocked gas (\citealt{Priestley22}). 
As for the density, for our purposes it is convenient to assume an average value, i.e.  $n_{\rm ISM}=1$ cm$^{-3}$ \citep{Gioannini17b},
corresponding to the average interstellar density of the Galaxy \citep{Ferriere98}. 
The prescriptions adopted for $M_{\rm clear}$ are summarised in the right panel of Fig. \ref{fig:dustpresc}.

\begin{figure}
    \centering
    \includegraphics[width=0.95\columnwidth]{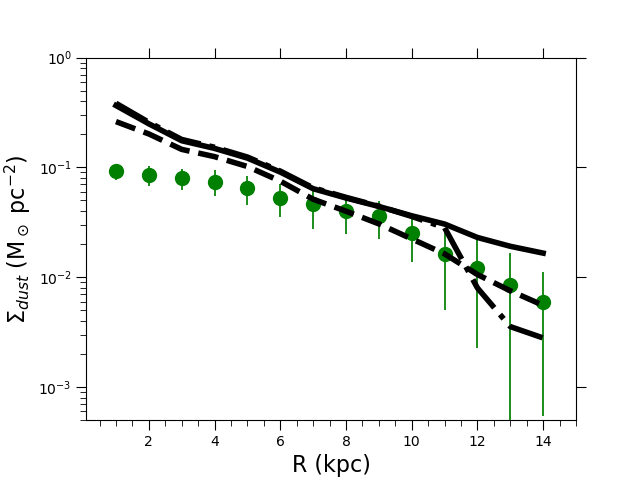}
    \includegraphics[width=0.95\columnwidth]{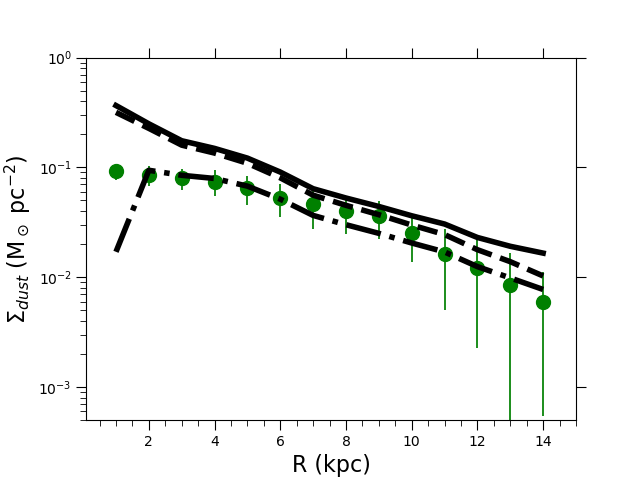}
    \caption{$\Sigma_{\rm dust}$ as function of galactocentric radius observed in M74 (green circles) and computed  
    adopting the prescriptions of \citet{Asano13} for dust accretion and  
    testing different timescales for dust destruction. 
    The top panel shows the effects of three different choices for the adopted cold gas fraction $X_{\rm cold}$, i.e. 
     $X_{\rm cold}=0.05$ (black dashed line, Model 1 of table \ref{tab_models}),
 $X_{\rm cold}=0.25$ (black solid line, Model 2),
 and a variable $X_{\rm cold}$ (Model 3) at different radii and computed as
    described in Sect.~\ref{sec_asa} (dash-dotted line).
    In the bottom panel, the results of a model with $X_{\rm cold}=0.25$ and different prescriptions for the cleared mass $M_{\rm clear}$ are
    shown, i.e. the one from \citet{Asano13} (black solid line, Model 2), \citet{Priestley22} (dashed line, Model 4),  and \citet{Temim15} (dash-dotted line, model 5).}
    \label{fig:dust_asano}
\end{figure}

As visible in the upper panel of Fig. ~\ref{fig:dust_asano}, all these assumptions lead to similar results,
i.e. an overestimate of the dust density profile at small radii, 
i.e. at $R<7$ kpc, with noticeable differences only in the outermost regions, where the metallicity
is lower than in the centre, and therefore the dust destruction timescale is the smallest in the whole galaxy. 

The three assumptions for the cold mass fraction 
imply a factor $\sim~3$ overestimation of the overall dust budget detected in M74. 

We now focus on one single model for dust accretion, 
where for the cold gas fraction we assume the 'cosmic' average value of $X_{\rm cold}=0.25$ and 
test different assumptions regarding destruction, by adopting various prescriptions for
the cleared mass $M_{\rm clear}$. 

In the recent work of \citet{Priestley22}, the effective mass cleared of dust is expressed as:
\begin{equation}
    M_{\rm clear}=\frac{8143}{1 + Z/(0.14 Z_\odot)} [{\rm M}_\odot].
    \label{eq:Priestley}
\end{equation}
We note that in Eq. \eqref{eq:Priestley}, the dependence of $M_{\rm clear}$ on the metallicity is stronger than as prescribed by \cite{Asano13}
(see the right panel of Fig. \ref{fig:dustpresc}). 
According to \citet{Priestley22}, Eq.~\ref{eq:Priestley} might overestimate the metallicity dependence of $M_{\rm clear}$ 
by neglecting kinetic sputtering, at variance with other authors that include this process \citep{Yamasawa11}. 
The results obtained adopting Eq. \eqref{eq:Priestley} (Model 4) are shown in the top panel of Fig. ~\ref{fig:dust_asano} as dashed lines. 

The adoption of these prescriptions 
do not lead to any significant improvement, as the observed inner dust profile is still overestimated. 
This is the effect of a significant metal production occurring in the innermost regions, that leads to 
a reduced vale of $M_{\rm clear}$ and therefore inefficient dust destruction.  
The large dust densities obtained at the innermost radii thus indicate that the models underestimate dust destruction, i.e. 
that the prescriptions adopted so far for $M_{\rm clear}$ are inadequate. To solve this conundrum, in the following we will test another possibility.  

The work of \citet{Temim15} includes the analysis of more than 80 SN remnants (SNR) in the Magellanic Clouds, in which the 
destroyed dust masses were determined by means of multiwavelength observations. In particular,
the sizes of the SNR were determined by means of X-ray data, whereas IR spectra obtained with Herschel enabled to estimate of the dust mass by
means of two-component SED fitting, including silicate and carbon dust. 
For the total effective swept-up mass of dust, \citet{Temim15} suggest a weak dependence on ISM density.
This weak dependence on the ISM density is due to the fact that the dust mass destroyed by SNR is mostly determined by 
the dust-to-gas value in the pre-shocked ISM. 
The typical values for the cleared (or swept-up) mass values derived by \citet{Temim15}
for carbon dust and silicates are of the order of $\sim 2000 ~M_{\rm \odot}$. 
However, in the pre-shocked ISM around the studied SNR, \citet{Temim15} measure dust-to-gas (DGR) ratios that are 
a factor of two to three higher than the average DGR galactic values.  
Considering that ${G}_{i} =  {G}_{i,\rm d}/DGR$, 
in our equations (e. g., eq. 10) we take into account an enhanced DGR around SN remnants by assuming a factor two larger cleared mass. 
\citet{Temim15} also suggest a weak dependence on metallicity of the swept-up mass, $M_{\rm clear} \propto \zeta_m^{-0.15}$, where $\zeta_m$ is the
metallicity normalized to its solar value, and taken to be equal to 0.3 for the Magellanic Clouds. 

Based on these results, we examine the effects of a constant value for $M_{\rm clear}$, i.e. independent on both density and metallicity,  
and assume $M_{\rm clear}=4000 M_{\rm \odot}$ (Model 5), therefore larger than the average cleared dust mass derived in SNR by \citet{Temim15}.  
The results of this assumption are shown by the dash-dotted profile in the bottom panel of Fig. ~\ref{fig:dust_asano}, that 
outline the significant improvement in accounting for the observed profiles with respect to the previous prescriptions at all radii,
but in the innermost regions, i.e. at $R<2$ kpc. 
This is due to the larger SFR, and therefore CC SN rate in the inner annulus ($R\sim1$ kpc),  by a factor $\sim$ 3 larger than
the values at 2 kpc.  

\begin{figure}
    \centering
    \includegraphics[width=0.95\columnwidth]{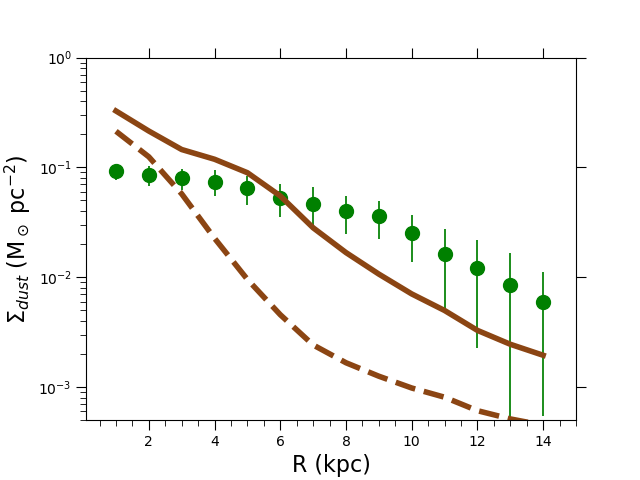}
    \includegraphics[width=0.95\columnwidth]{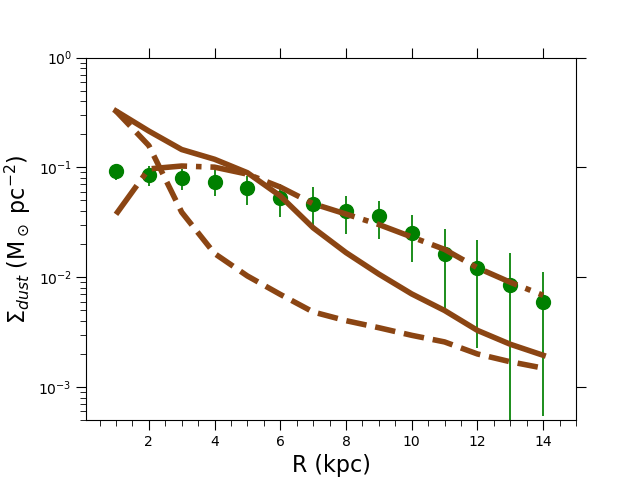}
    \caption{$\Sigma_{\rm dust}$ as function of galactocentric radius observed in M74 (green circles) and computed  
    adopting the prescriptions of \citet{Mattsson12} for dust accretion and  
    testing different timescales for dust destruction.
    In the top panel, we show the results obtained assuming the \citet{Asano13} (Model 6, brown solid line) and \citet{Priestley22} (Model 7, brown dashed line)
    prescriptions for the cleared mass $M_{\rm clear}$ and a constant value (500) for the parameter $\xi$ of Eq.~\ref{eq_tau0_mat12}. 
    In the bottom panel, we show the effects of different choices for the parameter $\xi$ as a function of the SF efficiency $\nu$. 
    Brown solid line: constant value for $\xi$ ($\xi=500$, Model 6); brown dashed line: $\xi=225~\nu$ (Model 8); brown dash-dotted line:   
    $\xi=150/\nu$ (Model 9). In each model, the adopted $M_{\rm clear}$ is from \citet{Asano13}. The solid green circles are as in Fig.  \ref{fig_tmap}.}
    \label{fig:dust_mattsson}
\end{figure}

\begin{figure}
 \includegraphics[width=0.95\columnwidth]{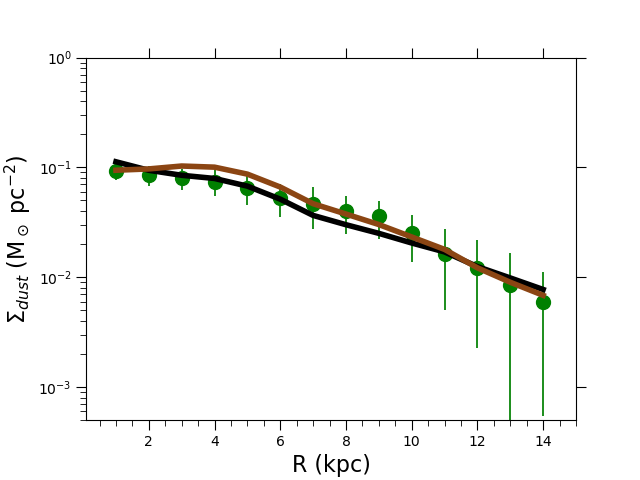}
  \caption{Best models obtained assuming the \citet{Asano13} (black solid line, Model 5' of Tab.\ref{tab_models}) and \citet{Mattsson12}
  (brown solid line, Model 9' of Tab.\ref{tab_models}) prescriptions for dust accretion. In both models, the adopted  $M_{\rm clear}$
  has been reduced by a factor of 1/3 in the innermost regions (R$<2$ kpc) to reproduce the dust mass density observed at $R=1$ kpc in the M74 galaxy. The solid green circles are as in Fig. \ref{fig_tmap}.}
  \label{fig:dust_best}
\end{figure}

\subsubsection{Mattsson et al. (2012) prescriptions for accretion}
\label{sec_mat}
\cite{Dwek98} expressed the rate per unit volume at which the dust grains grow by accretion of metals for the element $A$ 
\begin{equation}
\frac{dN_{\rm A}}{dt}= \alpha \pi a^2 n_{\rm g} (A) n_{\rm gr} \overline{\rm v},
\label{eq_rate}
\end{equation}
where  $\alpha$ is the sticking coefficient of element A to the dust grain,  $n_{\rm g}$ and
$n_{\rm gr}$ are the number density of gas particles and dust grains, respectively, and 
$\overline{\rm v} = \frac{8 k T}{\pi m_{\rm A}}$ is the mean thermal speed of atoms with mass $m_{\rm A}$ in interstellar gas 
with temperature $T$. 
By expressing Eq. \ref{eq_rate} in terms of dust, cold gas and metal surface densities, \cite{Mattsson12} showed that $\tau_0$ can be written as:
\begin{equation}
    \tau_0= \frac{\Sigma_{\rm gas}(t)}{\xi\, Z\, \dot{\Sigma}_{\rm *}(t)},
    \label{eq_tau0_mat12} 
\end{equation} 
where $\Sigma_{\rm gas}$ is the gas surface density, $\dot{\Sigma}_{\rm *}(t)$ is the SFR density and $\xi$ is a free parameter,
whose expected value is of the order of a few hundred, required to obtain $\tau_0\sim 10^7$ yr for a solar metallicity
and values of $\Sigma_{\rm gas}$ and $\dot{\Sigma}_{\rm *}(t)$ observed in the solar neighbourhood. 
In this case, it is first convenient to adopt a fixed value of $\xi=500$  and test various assumptions for $M_{\rm clear}$.
The results are shown in the top panel of Fig.~\ref{fig:dust_mattsson},
where we see that the adoption of both the \cite{Asano13} and \cite{Priestley22} prescriptions
for $M_{\rm clear}$ produce too steep dust mass profiles with respect to observations. 
We have tested also the adoption of the \cite{Temim15} prescriptions for $M_{\rm clear}$ (not shown in the plot),
finding no substantial improvement in reproducing the observed profile. 

One possible way to reduce the slope in the predicted dust profile is to adopt a variable $\xi$ parameter as a function of galactocentric distance.
To this aim, we will test a couple of different parametrisations in which $\xi$ will be expressed as a function of the SF efficiency $\nu$,
represented by a decreasing function of radius as shown in Fig.~\ref{fig_laws}. 
This will require sticking to one particular prescription for destruction, for which we adopt the \cite{Asano13} formula for $M_{\rm clear}$ (Eq. ~\ref{eq_mclear_asa13} ). 

In our first attempt, we assume $\xi$ as a linear function of the SF efficiency $\nu$: 
\begin{equation}
\xi = K \nu
\end{equation}
For the proportionality factor $K$, in the bottom panel of Fig.~\ref{fig:dust_mattsson} we show the results obtained assuming $K=225$ (Model 7). However,
the adoption of other values within the same order of magnitude does not lead to significantly different results.
This model is represented by the dashed line, showing a flatter profile at outer radii ($R>4$ kpc) than the model with constant $\xi$ (solid line), 
resulting from a decreased accretion rate (due to longer SF timescales) in the outer regions, in strong disagreement
with the observed profile. 

Conversely, the assumption of $\xi$ inversely proportional to $\nu$
\begin{equation}
\xi =  K'/\nu
\label{eq_matts_ok}
\end{equation} 
leads to reproduce satisfactorily the slope of the observed profile, as shown by the dash-dotted line in the bottom panel of Fig.~\ref{fig:dust_mattsson}.
In this case, the proportionality constant of Eq.~\ref{eq_matts_ok} is set to K'=150, a value optimised to reproduce also the normalisation of the observed
profile, that in this case is the dust surface density at R=2 kpc. 
As in the case of the best model obtained with the \cite{Asano13} prescriptions for dust accretion, the latter assumption does not allow us to 
account for the dust density in the innermost regions, i.e. at R=1 kpc, where the amount of dust is underestimated. 
This is due to a very high destruction rate in the innermost regions, that stems from the high star formation rate, thus implying a too high 
SN rate. 

Fig.~\ref{fig:dust_best} shows the results of the two best models 5' and 9', in which the dust density is reproduced at all radii, including also 
the inner region (R=1 kpc). In both models, the dust destruction rate has been reduced by a factor of 1/3 to increase the final
dust mass. Although in our model this assumption is performed ad-hoc and on an empirical basis, it has interesting implications and motivated
physical grounds, discussed in Sect.~\ref{sec:disc}. 

\begin{figure*}
    \centering
    \includegraphics[width=0.475\textwidth]{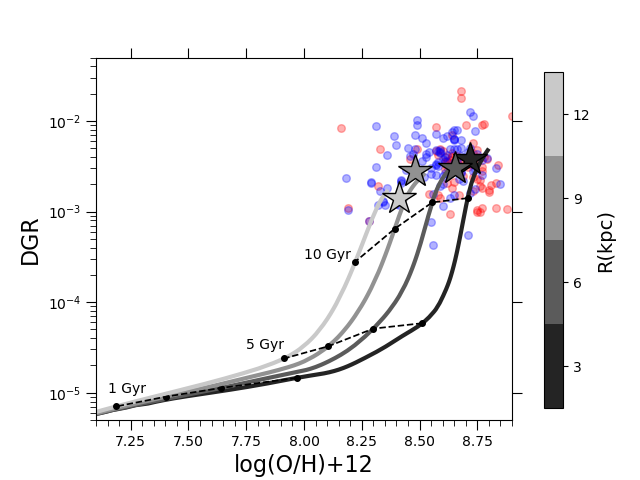}
    \includegraphics[width=0.475\textwidth]{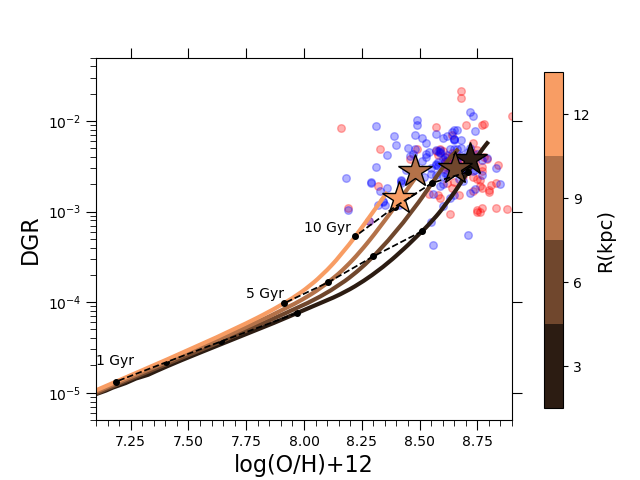}
    \caption{Dust-to-gas ratio (DGR) as a function of log(O/H)+12 for the best models at different galactocentric radii.
    Left panel: best model obtained assuming the \citet{Asano13} dust accretion prescriptions (Model 5' of Tab~\ref{tab_models}).
    The black, dark grey, grey and light grey curves represent the evolutionary tracks at radii 3 kpc, 6 kpc, 9 kpc and 12 kpc, respectively. 
    The stars represent the values measured in M74. 
    The red and blue circles are the DGR values observed in early-type (red) and late-type (blue) spiral galaxies of the DustPedia sample.
    The dashed lines are isochrones at 1 Gyr, 5 Gyr and 10 Gyr. 
    Right panel: best model obtained assuming the \citet{Mattsson12} dust accretion prescriptions (Model 9' of Tab~\ref{tab_models}.
    The dark brown, brown, light brown and orange curves represent the evolutionary tracks at radii 3 kpc, 6 kpc, 9 kpc and 12 kpc, respectively. 
    The other symbols and lines are as in the left panel.}
    \label{fig:scaling_DGR}
\end{figure*}

\begin{figure*}
    \centering
    \includegraphics[width=0.475\textwidth]{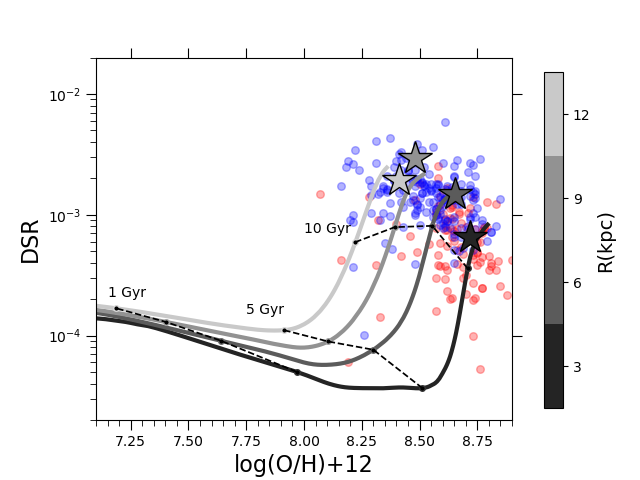}
    \includegraphics[width=0.475\textwidth]{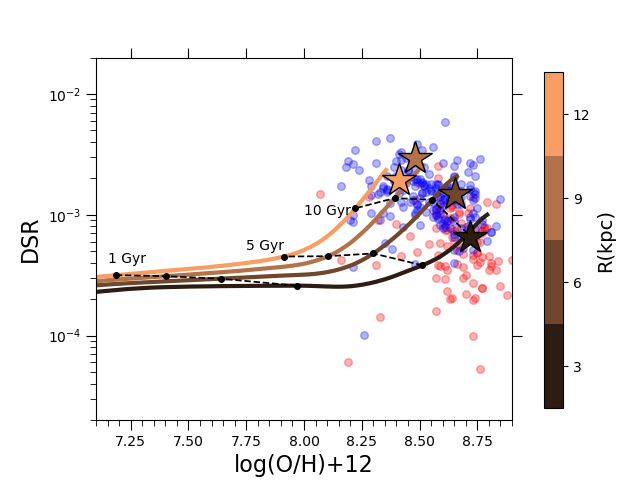}
    \caption{Dust-to-star ratio (DSR) as function of log(O/H)+12 for the best models at different galactocentric radii.
    Left panel: best model obtained assuming the \citet{Asano13} dust accretion prescriptions (Model 5' of Tab~\ref{tab_models}). 
    The black, dark grey, grey and light grey curves represent the evolutionary tracks at radii 3 kpc, 6 kpc, 9 kpc and 12 kpc, respectively. 
    Other symbols are as in Fig. ~\ref{fig:scaling_DGR}.
    Right panel: best model obtained assuming the \citet{Mattsson12} dust accretion prescriptions (Model 9' of Tab~\ref{tab_models}).
    The dark brown, brown, light brown and orange curves represent the evolutionary tracks at radii 3 kpc, 6 kpc, 9 kpc and 12 kpc, respectively. 
    The other symbols and lines are as in the left panel.}
    \label{fig:scaling_DSR}
\end{figure*}

\begin{figure*}
    \centering
    \includegraphics[width=0.475\textwidth]{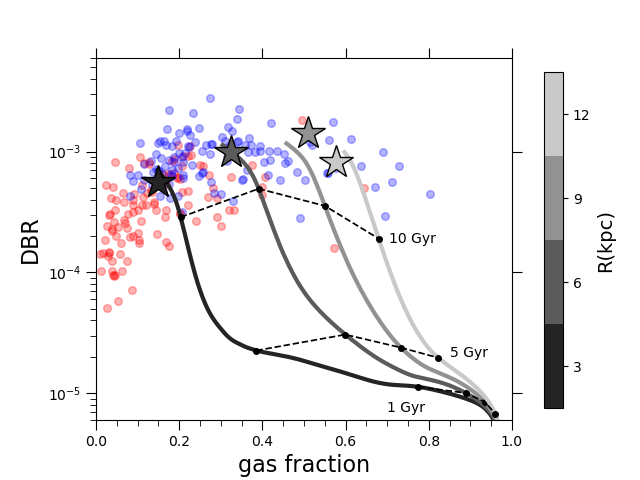}
    \includegraphics[width=0.475\textwidth]{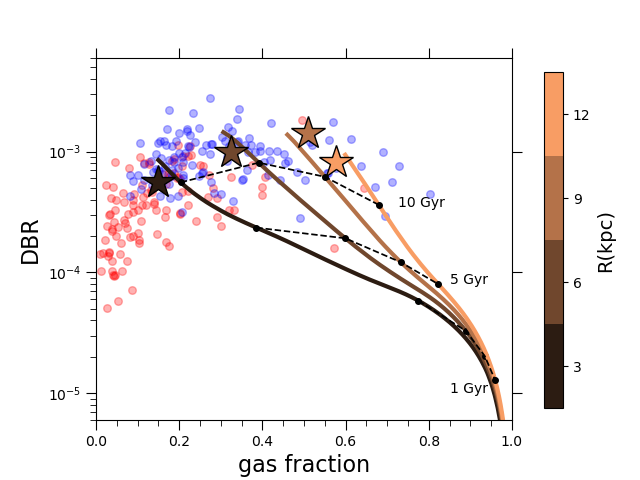}
    \caption{Dust-to-baryon ratio (DBR) as function of the gas fraction for the best models at different galactocentric radii.
    Left panel: best model obtained assuming the \citet{Asano13} dust accretion prescriptions (Model 5' of Tab~\ref{tab_models}).
    The black, dark grey, grey and light grey curves represent the evolutionary tracks at radii 3 kpc, 6 kpc, 9 kpc and 12 kpc, respectively.
    Right panel: best model obtained assuming the \citet{Mattsson12} dust accretion prescriptions (Model 9' of Tab~\ref{tab_models}).
    The dark brown, brown, light brown and orange curves represent the evolutionary tracks at radii 3 kpc, 6 kpc, 9 kpc and 12 kpc, respectively.  
    The other symbols and lines are as in Fig. \ref{fig:scaling_DGR}.}
    \label{fig:scaling_DBR}
\end{figure*}

\subsubsection{Dust-to-gas and Dust-to-stellar ratios}
By construction, the final DGR and dust-to-stellar (DSR) ratios obtained with our best models 
match the values observed in M74. 
Here we discuss the predicted DGR and DSR as a function of metallicity, galactocentric distance
and compare our results with the integrated values observed in other DustPedia galaxies. 
The sample we consider in our analysis includes a sub-set of late- and early-type spiral galaxies for which the metallicity (expressed as 12+log(O/H)),
total dust, gas and stellar mass have been measured and is composed of
393 galaxies in total (163 early-type and 230 late-type spirals). 

For details on how the measurements were performed and the data collected, we refer the reader to \cite{Casasola17} and \cite{Casasola20}.
The DGR as a function of metallicity is presented in Fig. \ref{fig:scaling_DGR}, in which the results of Model 5' (9') is reported in the left (right) panel.
In both cases, the model values computed at the final time  are in excellent agreement with the data for the sub-sample of late-type galaxies. 
 The curves of different colors describe the evolution at 4 representative radii,
from 3 kpc to 12 kpc. Isochrones have been plotted for both model sets, computed at three different evolutionary times (1 Gyr, 5 Gyr and 10 Gyr, 
dashed lines), useful
to examine the evolution of the DGR at different radii.
As for Model 5', the slope of the isochrones at the outer radii (R$>8$ kpc) progressively steepen with time, indicating
that also the DGR gradient steepens appreciably at late evolutionary stages ($\gtrsim$ 10 Gyr). 
This occurs because of the dependence of the accretion timescale $\tau_{\rm g}$ on metallicity (Eq.~\ref{eq_asa_tg} ), leading to 
longer timescales in more external regions as due to longer infall and SF timescales. 

Less differences are visible in the evolution of the DGR at R$<6$ kpc, characterised by 
a flatter behaviour of the isochrones at times $\le 10$ Gyr and a flatter gradient at the final time than the outer regions.

At the final time, the track computed at the inner radius overlaps with the DGR observed in early-type spirals, presenting values comparable to late-type
galaxies but characterised by larger metallicities. 
The tracks of Model 9' (right panel of Fig. \ref{fig:scaling_DGR}) are steeper than Model 5' and show a stronger evolution, visible in the larger
DGR ratios at various radii at 1 and 5 Gyr. 

The total DSR of the models is $\sim 0.001$, in good agreement with the values measured in late spirals of the Dustpedia sample. 
The evolution of the DSR at different galactic radii is shown in Fig. \ref{fig:scaling_DSR}, in which the results of models 5' and 9' are reported in
the left and right panels, respectively, compared to the local values observed in DustPedia galaxies (open circles).
In this case, the behaviour of the isochrones is the inverse of the DGR, in that they steepen at lower radii,
specifically decreasing as a function of radius, remaining approximately flat at R$>6$ kpc. 
This is due to the shape of the stellar mass profile, particularly steep at small radii ($<$3 kpc),
where the dust mass profile is essentially flat (see Fig.~\ref{fig_tmap}).
The final values of the model at different radii are in good agreement with the values observed in late-type DustPedia galaxies.
The end-point of the track at R=2 kpc overlaps with the DSR of the lowest mass early type spirals, presenting DSR$\gtrsim 3 \times 10^{-4}$. 

The evolution of Model 9' at various radii is different than the one of Model 5', in that the tracks of the former present
a flatter early evolution with metallicity and they evolve more closely, presenting a singnificantly smaller range of DSR at 5 Gyr. 
At later times, the two models show little differences. 
The closer evolution of the tracks of Model 5' and 9' at late evolutionary times are visible also in Fig. \ref{fig:scaling_DBR},
in which we show the dust-to-baryon ratio computed at various radii as a function of the gas fraction.
The analysis of this quantity outlines the steeper growth of the tracks of Model 9', that at 5 Gyr show DBR$>\gtrsim 1 \times 10^{-4}$,
whereas at the same time, Model 5' shows still DBR values of a few $\times 10^{-5}$ at all radii. 

Moreover, this plot emphasises the separation between DustPedia late- and early-types, 
with the latter presenting in most cases lower gas fractions ($\lesssim 0.2$) than the former. In their later stages,
i.e. at evolutionary times $>10$ Gyr, the theoretical tracks of the innermost radii of models 5' and 9' overlap with the
DBR of the local early type-galaxies characterised by larger gas fraction ($\sim 0.2$).

\subsection{Discussion}
\label{sec:disc}
In this work, we figured out simple recipes to model destruction and accretion
that can reproduce the observed profile of M74 without any ad-hoc assumption. 
This holds at all radii except in the innermost regions (R$<2$ kpc),
where we empirically decreased the dust cleared mass to match the observational data. 
In this section, first we discuss possible physical reasons for this choice and on the possible radial dependence
of dust destruction, as a result of a different impact of stellar feedback at different galactic radii. 

We are aware that the adopted recipes might not be unique; in this regard, our study is useful to outline 
the degeneracy between the main parameters regulating dust destruction and accretion. 
We also discuss the relative roles of these processes in our models, that are still a matter of 
debate, and compare our findings with other relevant works.

\subsubsection{On the efficiency of dust destruction at different galactic radii}
How the destruction of dust depends on the galactocentric radius and, more in general, on the stellar feedback in
different environments, is largely unknown. Although a few works dealt with the
destruction of dust from single SN (\citealt{Silvia10}, \citealt{Martinez-Gonzalez19})), 
the destruction of dust from multiple supernovae has rarely been treated. 
In a multiphase ISM, it is known that the destruction is related to the mass exchange between the 
hot and cold media (\citealt{McKee89,Peters17}). 
Such exchange determines the circulation of gas from the warm neutral medium,
where dust abundances are reduced by destruction in interstellar shocks 
generated by massive stellar feedback, 
and matter enriched with dust due to accretion on grain surfaces at higher gas densities. 
In this context, the knowledge of the filling factor of the warm and hot medium driven by pre-SNe and SN feedback 
 and its dependence on the environment is fundamental. Such quantity is determined by 
mutual interactions of SN remnants at various radii and their impact on the multiphase medium, that 
has instead already been treated in various works. 

In the innermost regions of galactic discs, the evolution of SN remnants might indeed be different than
at outer radii. 
\cite{Yalinewich17} showed that at small distances from the Galactic Centre, SN remnants evolve differently than at outer radii
because of the higher abundance of young massive stars.
These stars continuosly emit stellar winds with total mass loss rate of $\sim 10^{-3} M_{\rm \odot}$/yr 
and propagating with typical velocity of $\sim 700$ km/s \citep{Rockefeller04}.   
Instead of a static, 
uniform density medium, a SN exploding near the Galactic center  propagates into a continuously wind-swept environment and 
a steep density profile, causing a non-spherical expansion of the SN remnant \cite{Rimoldi15}, due to
a faster deceleration of the fluid moving towards the centre.
This might influence the interactions between different SN bubbles and the impact of stellar feedback.

By means of three-dimensional hydrodynamic simulations of the galactic disc, 
\cite{Melioli09} studied the evolution of galactic fountains generated by stellar OB associations
at various galactocentric distances, in a model that includes 
differential galactic rotation and a gaseous halo.
The energy restored by OB associations generates several hot cavities in the disc, whose
lifetime and overlapping are determined by the duration of SN activity, local density and rotation.
These factors strongly influence the filling factor of the hot cavities, that drops sharply towards the smallest galactic radii.
This finding is supported also by observations of local galactic discs.
In  a detailed study of the HI holes in the spiral galaxy NGC 69456, 
\cite{Boomsma08} find that the hole covering factor drops sharply towards smallest galactic
radii. This effect is ascribed to both a drop of the H I column density
and to the stronger shear, shortening the hole lifetimes and therefore decreasing their overlapping. 
\cite{DellaBruna22}  present a study based on the Multi Unit Spectroscopic Explorer (MUSE) of the
properties of HII regions in the nearby spiral galaxy M83, 
aimed at quantifying the effects of stellar feedback in different galactic environments.
They compare the internal pressure of HII regions with the environmental pressure,
finding that in the innermost regions (R$\le$ 0.5 kpc) the latter is almost one order of magnitude higher than the former,
indicating that stellar feedback is not sufficient to drive the expansion of the ionised regions.
On the other hand, in more external zones, HII regions are over-pressurized, therefore they can expand.
This indicates that the strength of early stellar feedback increases at increasing  galactic radii.

In this framework, in our model a particularly inefficient destruction at galactic radii $<1$ kpc
might be qualitatively explained by a lesser impact of stellar feedback at small galactocentric distances. 
To test this hypothesis quantitatively, 3D hydrodynamical simulations of a realistic galactic
disc will be required, in which stellar feedback and dust destruction are taken into account.

\subsubsection{On the balance between destruction and accretion in galaxies}
\begin{figure}
    \centering
    \includegraphics[width=1\columnwidth]{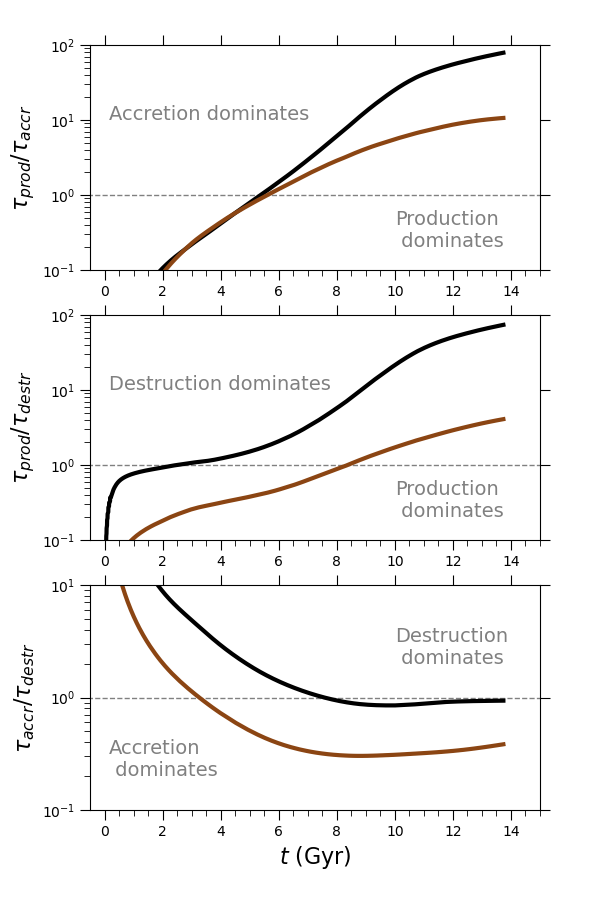}
    \caption{Ratios between accretion, destruction and stardust production timescales
    for Model 5' (black lines) and Model 9' (brown lines) of Tab~\ref{tab_models} as a function of time and at a galactic radius of 8 kpc.
    Top panel: ratio between stardust production and accretion timescales.
    Central panel: ratio between stardust production and destruction timescales.
    Bottom panel: ratio between accretion and destruction timescales.
    In each panel, the thin, grey dashed line indicates a ratio equal to 1, corresponding to the equilibrium between the two processes.} 
    \label{fig:ratio_tau}
\end{figure}

One often debated aspect in studies of dust evolution concerns which
processes among stardust production, accretion and destruction affect most the dust budget in galaxies. 
In Fig. \ref{fig:ratio_tau}, we show the relative roles of these mechanisms in the best models of our set,
i.e. Model 5' and Model 9', through the ratio between the timescales of each processes, both computed
at a galactocentric radius $R=8$ kpc, representative of the entire galaxy. 
The accretion and destruction timescales are defined in Sect.~\ref{s:dust_model},
whereas the production timescale
is the ratio between the cumulative mass of dust 
 and the total stardust production rate, computed including both AGB stars and SNe. 
In Model 5', the stardust production dominates over accretion and destruction only
at early times, i.e. up to $\sim 5$ Gyr and 2 Gyr, respectively. 
Model 9' has a more prolonged dominion of production by stars, extending
up to 6 Gyr and 8 Gyr over accretion and destruction, respectively.
Model 5' settles on a quasi-equilibrium between destruction and accretion after
8 Gyr, that continues up to the present time.
At a comparable epoch the ratio between the timescales of these processes becomes approximately
constant in Model 9', that also protracts to the present time. 
However, the two models predict two distinct behaviors 
regarding the relative roles of dust production and destruction and their evolution, as 
they settle onto different values of the ratio $\tau_{\rm acc}/\tau_{\rm destr}$. 

Several models rely on the primary role of accretion in determining the galactic dust budget
(e. g., \citealt{Dwek98,Calura08,Gall11,Valiante11,Mattsson14}).   
On the other hand, \cite{Gall18} performed an empirical study of the dust mass in high- and low-$z$ galaxies,
in which they explore a simple dust production scenario that applies to star-forming systems.
They considered star-forming massive galaxies  and local galaxies, including the Milky Way and the Magellanic Clouds,
and a wide set of measurements of the dust mass in SNe and SN remnants at various stages. 
In their study, the observed dust masses can be
accounted for by a simple model in which SNe produce the
majority of the dust  in the current star formation episode, on timescales $\lesssim$ 10-100 Myr. 
However, their results do not exclude the formation of dust through rapid grain growth out of the metals
expelled by massive stars. Rapid destruction and
reformation of the dust is also allowed, as long as the
two terms are roughly balanced, similar to Model 5'. 
At the moment it is very difficult to determine which of these possibilities is the most realistic.
In fact, the study of the cosmic evolution of the comoving mass density of dust indicates that the destruction
must play an important role. This is indicated by the decrease of the dust mass at $z <2$ \citep{Pozzi20},
mainly reproduced by models in which 
the observed decrease of the dust mass is determined by destruction and
stardust production has a marginal role at these epochs \citep{Gioannini17b}.
In the framework of our models, in this case the implication is that accretion also must play a primary role in the survival of dust in galaxies.

%%%%%%%%%%%%%%%%%%%%%%%%%%%%%%%%%%%%%%%%%%%%%%%%%%%%%%%%%%%
\section{Conclusions} \label{sec4}
The DustPedia project offers a wealth of multi-wavelength constraints
on a large sample of local galaxies of various morphological types \citep{Davies17,Davies19,Casasola20}.
In the past, these constraints have been used within a statistical framework,
to achieve constraints on chemical and dust evolution parameters
for nearby galaxies \citep{DeVis21}. 

In this paper, we introduce a multi-zone chemical evolution model of 
one DustPedia galaxy, M74, 
calibrated by means of MCMC methods. 
For this purpose, we use exquisite, high-resolution data collected 
to study spatially-resolved fundamental scaling relations (such as the Main Sequence 
and the Schmidt-Kennicutt law) in disc galaxies \citep{Enia20,Morselli20,Casasola22}. 
We take into account the stellar mass and gas profiles obtained in these studies,
and use Bayesian analysis to constrain two fundamental parameters defining the evolution 
of M74, the accretion timescale $\tau$ and the SF efficiency $\nu$, in both cases
as a function of galactocentric radius $R$, defined as the distance between
a given annuli and the centre of M74. 
The best model derived with this method is then used to investigate the dust content of M74,
comparing the observed dust density profile with the results of our chemical
evolution model. Various prescriptions have been considered for
two key parameters, i.e. the typical dust accretion timescale $\tau_{0}$ and the
mass of gas cleared out of dust by a SN remnant, $M_{\rm clear}$, playing a primary role in 
regulating the dust growth and destruction rate, respectively. 
Our results can be summarised as follows.
\begin{enumerate}
\item Our analysis supports an accretion timescale increasing with $R$,
thus supporting an 'Inside-Out' formation for M74. 
This confirms the result found for the accretion timescale of the Milky Way considering
the spatial dependence of the chemical abundances obtained from various diagnostics, such as the 'G-dwarf' distribution
at various galactocentric radii and the abundance gradient, as traced by the gas or  young stars \citep{Matteucci89, Romano00, Chiappini01}. 

The SF efficiency shows a decreasing trend as a function of radius. 
This is also consistent with what was found in the Milky Way from the analysis of the observed abundances for various chemical elements
(\citealt{Grisoni18}, \citealt{Palla20a}). 

For both quantities, our analysis indicates a weaker radial dependence than in the Milky Way.
This implies lesser variations of the gas accretion and SF rates with radius, that could manifest in a
lower dispersion of the stellar ages or chemical abundaces as a function of radius. 
\item We have tested the effects of radial gas flows on the observables considered
in this work, i.e. the stellar and gas density profiles and the metallicity gradient.
We have considered various cases for the most critical quantity characterising this process, i.e. 
the radial velocity pattern, including constant and radial-dependent  values,
by simultaneously  considering other assumptions for the radial dependence of the infall and SF law.
We have found that even values as low as -0.5 km/s are enough to steepen significantly the metallicity gradient and the gas
density profile.
By assuming net radial inflows in the galaxy and by
varying the prescriptions on the radial dependence of gas accretion and SF, 
we are not able to reproduce simultaneously the observational constraints of M74, 
i.e. the stellar density, gas density and metallicity gradient.
\item Considering various functional forms for the characteristic dust accretion timescale and the cleared dust mass,
we are able to account for the dust mass profiles observed in M74.
Our study outlines the degeneracy between the most fundamental accretion and destruction parameters in shaping the
interstellar dust content in galaxies, as two models with different prescriptions provide remarkably 
similar results for dust density profile.
These two models envisage a different current balance between destruction and accretion, in that both an equilibrium and
a dominion of accretion over destruction can equally reproduce the available constraints.
On much lager scales, a scenario in which destruction plays a marginal role seems perhaps excluded by the observed
decreasing evolution of the comoving dust mass density, difficult to account for by means of astration only. 
\item Despite their degeneracy, both the two best models obtained here can reproduce the observed
dust density profile at all radii except in the innermost regions (R$<2$ kpc),
where the dust content is underestimated.
In both cases, the dust destruction rate has to be reduced by a factor of 1/3 to increase the dust mass observed in
the most central regions of M74.
A particularly inefficient destruction at small galactic radii $<1$ kpc
can be qualitatively explained by a lesser impact of stellar feedback at small galactocentric distances.
This is supported by studies performed with
hydrodynamical simulations of a MW-analog, showing that filling factor of the hot cavities generated by OB stars
drops sharply towards the smallest galactic radii \citep{Melioli09} and is confirmed also by observational studies of
the spatial distribution of HII regions in local discs \citep{DellaBruna22}.  
\end{enumerate}

Although the results presented in this work can
be improved in various aspects, our aim is to deliver a model that accounts for the star formation history of the Dustpedia galaxy M74, whose reliability has been assessed on a quantitative basis, using a parametric approach and considering a pre-determined set of observational constraints.
This aim is complementary to the ones of other models, such as cosmological ones, in which it is more difficult to tailor a model to a specific system that accurately
accounts for observables such as gas and stellar density profiles. 
In the future, we plan to use the methods described in this paper to
extend our study to more DustPedia galaxies, in order to derive fundamental
constraints on their star formation history and the evolution of their dust content. 
This will lead to significant progress in our understanding of galactic evolution,  
outlining further the roles played by the most fundamental parameters that regulate it and 
 the differences between individual galaxies.

\section*{Data Availability}
The data underlying this article are available from the 
corresponding author, upon reasonable request.

%%%%%%%%%%%%%%%%%%%%%%%%%%%%%%%%%%%%%%%%%%%%%%%%%%%%%%%%%%%
%%%%%%%%%%%%%%%%%%%%%%%%%%%%%%%%%%%%%%%%%%%%%%%%%%%%%%%%%%%
%%%%%%%%%%%%%%%%%%%%%%%%%%%%%%%%%%%%%%%%%%%%%%%%%%%%%%%%%%%
%%%%%%%%%%%%%%%%%%%%%%%%%%%%%%%%%%%%%%%%%%%%%%%%%%%%%%%%%%%
\section*{Acknowledgements}
An anonymous referee is acknowledged for useful suggestions.
We acknowledge financial support from grant PRIN MIUR 20173ML3WW$\_$001. 
FC also acknowledges support from PRIN INAF 1.05.01.85.01. 
MP acknowledges funding support from ERC starting grant 851622 DustOrigin and
computing resources from the “accordo quadro MoU per lo svolgimento di attivit\'a congiunta di ricerca nuove frontiere in astrofsica: HPC e data exploration di nuova generazione”
between CINECA and INAF. 
MP also thanks Ilse De Looze, Monica Relaño and Stefan Van der Giessen for the fruitful discussions. 
VC, SB and FP acknowledge funding from the INAF Main Stream 2018 program “Gas-DustPedia: A definitive view of the ISM in the Local Universe”. 
FC, VC, SB and FP acknowledge funding from the INAF Mini Grant 2022 program “Face-to-Face with the Local Universe: ISM's Empowerment (LOCAL)”.
%%%%%%%%%%%%%%%%%%%%%%%%%%%%%%%%%%%%%%%%%%%%%%%%%%



\bibliographystyle{mnras}
\bibliography{mndustpedia_rev} % if your bibtex file is called example.bib

%\begin{thebibliography}{99}
%\bibitem[]{bel08}
%\bibitem[Calura \& Menci(2009)]{calura2009} Calura, F., \& Menci, N.\ 2009, \mnras, 400, 1347 
%\bibitem[]{tes02}
%\end{thebibliography}

%%%%%%%%%%%%%%%%%%%%%%%%%%%%%%%%%%%%%%%%%%%%%%%%%%

%%%%%%%%%%%%%%%%% APPENDICES %%%%%%%%%%%%%%%%%%%%%

%\appendix



%%%%%%%%%%%%%%%%%%%%%%%%%%%%%%%%%%%%%%%%%%%%%%%%%%

% Don't change these lines
\bsp	% typesetting comment
\label{lastpage}
\end{document}